\documentclass[twocolumn,english,prd,superscriptaddress,nofootinbib,preprintnumbers]{revtex4}

\makeatletter

\usepackage[T1]{fontenc}
\usepackage[latin1]{inputenc}
\usepackage{babel}
\usepackage{amsmath,amssymb}
\usepackage{dcolumn,tabularx}
\usepackage{natbib}
\usepackage{url}

\usepackage{ifpdf}
\ifpdf   
   \usepackage[pdftex]{graphicx}
   \DeclareGraphicsExtensions{.pdf,.png}
   \usepackage{pgf}
\else 
   \usepackage{graphics}
   \DeclareGraphicsExtensions{.eps,.png}   
\fi

\usepackage[hyperfootnotes=false,pdfstartview=,colorlinks=true,linkcolor=purple, citecolor=blue,urlcolor=blue]{hyperref}

\newcommand{\dd}{\textrm{d}}

\newcommand{\lcdm}{$\Lambda$CDM}
\newcommand{\nqso}{N_{\rm QSO}}
\newcommand{\zqso}{z_{\rm QSO}}
\newcommand{\lya}{Ly$\alpha$}
\newcommand{\sun}{\mbox{\tiny SUN}}
\newcommand{\unit}[1]{\ensuremath{\, \mathrm{#1}}}

\makeatother

\begin{document}

\title{Distinguishing Between Void Models and Dark Energy with Cosmic Parallax
and Redshift Drift}

\author{Miguel Quartin}
\email{m.quartin -.at.- thphys.uni-heidelberg.de}
\affiliation{Institut für Theoretische Physik, Universität Heidelberg, Philosophenweg
16, 69120 Heidelberg, Germany}

\author{Luca Amendola}
\email{l.amendola -.at.- thphys.uni-heidelberg.de}
\affiliation{Institut für Theoretische Physik, Universität Heidelberg, Philosophenweg
16, 69120 Heidelberg, Germany}
\affiliation{INAF/Osservatorio Astronomico di Roma, V. Frascati 33, 00040 Monteporzio
Catone, Roma, Italy}

\date{December 23, 2009}

\begin{abstract}
Two recently proposed techniques, involving the measurement of the
cosmic parallax and redshift drift, provide novel ways of directing
probing (over a time-span of several years) the background metric
of the universe and therefore shed light on the dark energy conundrum.
The former makes use of upcoming high-precision astrometry measurements
to either observe or put tight constraints on cosmological anisotropy
for off-center observers, while the latter employs high-precision
spectroscopy to give an independent test of the present acceleration
of the universe. In this paper, we show that both methods can break
the degeneracy between LTB void models and more traditional dark energy
theories. Using the near-future observational missions Gaia and CODEX
we show that this distinction might be made with high confidence levels
in the course of a decade.
\end{abstract}
\maketitle

\section{Introduction}

\label{sec:intro}

The enigma of the cosmic acceleration has solicited explanations that
range from new matter components with negative pressure, to modifications
of gravity, to large-scale violations of the cosmological principle
of homogeneity and isotropy. The latter class of models is probably
the most controversial but has the merit of linking explicitly the
acceleration (apparent or real) to the formation of non-linear structures
and of dispensing with unknown and so far unseen new cosmic components.

Any violation of the cosmological principle means that the simple
structure of the Friedmann-Robertson-Walker~(FRW) metric can no longer
be adopted, not even approximately, as a description of the universe
properties. The simplest possibility is to adopt in place of the FRW
metric the spherically symmetric structure of the Lemaître-Tolman-Bondi~(LTB)
metric, as suggested by various authors (e.g.~\cite{humphreys97,celerier00,alnes06a,garcia-bellido08a})
ever since the discovery of acceleration (a similar but non-LTB void model was also investigated in~\cite{tomita00}).
In order to reproduce the accelerated expansion, the LTB structure must allow for a faster expansion
inside than outside, which is generally (although not necessarily~\cite{celerier09})
obtained by a radial density profile that generate a huge ($\approx1-2$
Gpc) void. Notice that in this case the observed supernovae acceleration
is not real but rather due to the comparison of different sources
(inside and outside the void) and to the assumption of homogeneity;
in reality, in a LTB universe composed uniquely by dust matter there
is no real acceleration, except possibly (i.e., depending on the density
profile model) near the edge. Although a single huge LTB bubble with
the Milky Way right near the center is undoubtedly a contrived configuration,
this can be thought of as a first approximation towards a more realistic
model, for instance a collection of ellipsoidal voids and {}``meatballs''
of different sizes~\cite{biswas08,marra07,kainulainen09}. In any
case, almost all other dark energy models suffer from high-degrees
of fine-tuning, either in the necessary initial conditions or in the
form of the coincidence problem~\cite{amendola06,quartin08}.

As we discuss in more detail in the next section, the LTB metric allows
for two spatial degrees of freedom, that could be employed to reproduce
any line-of-sight expansion rate and any line-of-sight inhomogeneity.
In particular, LTB models (although not necessarily voids) can mimic the observed accelerated expansion
rate $H(z)$ and the observed source number counts at the same time~\cite{partovi84,celerier09}.
Because of this flexibility, and because of the isotropy with respect
to the center observer, ruling out the LTB model is not a trivial
task.

Although we sometimes take for granted that in cosmology we can only
access the surface of a single light cone, this is by no means true.
We can in fact receive CMB light scattered from distant sources, for
instance from the hot intra-cluster medium of galaxy clusters through
the Sunyaev-Zel'dovich effect, which comes from inside our light cone.
The spectrum of these scattered CMB photons will be distorted from
their original black-body spectrum and the amount of deviation is
proportional to the peculiar velocity of the cluster with respect
to the CMB scattering surface~\cite{garcia-bellido08b}. This effect
can be employed to map the cosmic peculiar velocity field and therefore
adds to the expansion rate and the number counts a third spatial function
that can break the fundamental degeneracy of LTB and FRW. Similarly,
since during reionization the CMB photons are scattered towards us
by structures that are located off-center, their spectrum will be
the sum of black-body spectra at different temperatures and therefore
will again deviate from a black-body spectrum~\cite{caldwell08}.
The amount of deviation depends on the distance with respect to the
center and provides again an additional piece of information that
can break the degeneracy.

In the above two examples one receives information from inside our
own light cone making use of sources along it. Two additional techniques
recently proposed explore instead the \emph{exterior} of our present
light cone by observing the same sources at two different instants
of time. In other words, by probing two or more different (albeit
very close) light cones.

The first method relies on high-precision spectroscopy. If the time
span $\Delta t$ is large enough, one can detect small changes $\Delta_{t}z$
in the source redshift proportional to the local expansion rate: this
is the so-called Sandage effect~\cite{sandage62} or redshift drift~\cite{loeb98,uzan08}.
As we will show below, the redshift drift can be used to distinguish
between real acceleration driven by dark energy ($\Delta_{t}z>0$)
and apparent acceleration ($\Delta_{t}z<0$). This technique has been
presented on a general basis in~\cite{partovi84,yoo08} but never discussed
in any detail nor compared to dark energy cosmologies.

The second method requires high-precision astrometry and exploits
the fact that off-center observers see an anisotropic space. We already
know that the distance from the LTB center is limited to less than
$50-100$ Mpc$/h$ by the observed isotropy of the CMB, of number
counts and of the supernovae Hubble diagram. It is however possible
to considerably reduce this upper limit by exploiting the recently
proposed cosmic parallax (CP) effect~\cite{quercellini09a,ding09,fontanini09,quercellini09b}.
The CP is the change in the angular separation of distant sources
induced by the differential expansion rate in anisotropic universes.
Any off-center observer in a LTB void will experience an anisotropic
expansion and therefore a CP, proportional (at first order) to the
distance from the void center. In~\cite{quercellini09a} this was
applied to voids and in~\cite{fontanini09,quercellini09b} to Bianchi
I models.

The redshift drift and the cosmic parallax form a new set of real-time
cosmic observables. In this paper we discuss both methods. In particular,
we calculate the former in the case of an LTB void and show that,
with the proposed EELT instrument CODEX~\cite{codex}, it is one
of the most promising way to distinguish voids from standard dark
energy models. For the cosmic parallax, we generalize and improve
on a number of points the previous treatments: we extend the analytical
estimates for sources at arbitrary positions, make a more accurate
estimate of the observational power of both Gaia~\cite{perryman01,bailer-jones05}
and the SIM Lite Astrometric Observatory~\cite{goullioud08,simbook}
missions using a realistic quasar distribution (taking into account
two major systematics), investigate the redshift dependence of both
signal and noise and propose a possible Figure of Merit for future
astrometry missions. We also include a third void model from the literature~\cite{garcia-bellido08a}
to better understand the model-dependance of both real-time cosmic
observables studied herein.

\section{LTB Void Models}

\label{sec:ltb}

The LTB metric can be written as (primes and dots refer to partial
space and time derivatives, respectively):
\begin{equation}
    {\textrm{d}}s^{2}=-{\textrm{d}}t^{2}+\frac{\left[R'(t,r)\right]^{2}}{1+\beta(r)} {\textrm{d}}r^{2}+R^{2}(t,r){\textrm{d}}\Omega^{2},\label{eq:LTB}
\end{equation}
 where $\beta(r)$ can be loosely thought as position dependent spatial
curvature term. Two distinct Hubble parameters corresponding to the
radial and perpendicular directions of expansion are defined as
\begin{align}
    H_{||}=\dot{R'}/R'\,,\label{eq:def-Hpar}\\
    H_{\perp}=\dot{R}/R\,.\label{eq:def-Hperp}
\end{align}
Note that in a FRW metric $R=ra(t)$ and $H_{||}=H_{\perp}$. The
Einstein equations for pressureless matter reduce to
\begin{align}
    H_{\perp}^{2}+2H_{||}H_{\perp}-\frac{\beta}{R^{2}}-\frac{\beta'}{RR'}= & \;8\pi G\rho_{m}\,,\label{eq:ltb-fri1}\\
    6\frac{\ddot{R}}{R}+2H_{\perp}^{2}-2\frac{\beta}{R^{2}}-2H_{||}H_{\perp}+\frac{\beta'}{RR'}= & -8\pi G\rho_{m}\,.\label{eq:ltb-fri2}
\end{align}
They can be further summed into a single equation which can be integrated
once to give the classical cycloid equations
\begin{equation}
    H_{\perp}^{2}=\frac{\alpha(r)}{R^{3}}+\frac{\beta(r)}{R^{2}}\,,\label{eq:hperpeq}
\end{equation}
where $\alpha(r)$ is a free function that we can use along with
$\beta(r)$ to describe the inhomogeneity. From this we can define
an effective density parameter $\Omega_{m0}(r)=\Omega_{m}(r,t_{0})$
today:
\begin{equation}
    \Omega_{m0}(r)\equiv\frac{\alpha(r)}{R_{0}^{3}H_{\perp,0}^{2}}\,,
\end{equation}
where $R_{0}\equiv R(r,t_{0}),\, H_{\perp,0}\equiv H_{\perp}^{2}(r,t_{0})$
and an effective spatial curvature
\begin{equation}
    \Omega_{K0}(r)=1-\Omega_{m0}(r)=\frac{\beta(r)}{R_{0}^{2}H_{\perp,0}^{2}}\,.\label{eq:ltb-curv}
\end{equation}
Note there another possible (and non-equivalent) definition is sometimes found in the literature~\cite{alnes06a}.
Eq.~(\ref{eq:hperpeq}) is the classical cycloid equation\index{cycloid equation}
whose solution for $\beta>0$ is given parametrically by
\begin{align}
    R(r,\eta) & \,=\,\frac{\alpha(r)}{2\beta(r)}(\cosh\eta-1)\nonumber \\
     & \,=\,\frac{R_{0}\Omega_{m0}(r)}{2[1-\Omega_{m0}(r)]}(\cosh\eta-1),\label{eq:sol-R}\\
    t(r,\eta)-t_{B}(r) & \,=\,\frac{\alpha(r)}{2\beta^{3/2}(r)}(\sinh\eta-\eta)=\nonumber \\
     & \,=\,\frac{\Omega_{m0}(r)}{2[1-\Omega_{m0}(r)]^{3/2}H_{\perp,0}}(\sinh\eta-\eta),\label{eq:sol-beta-t}
\end{align}
where the ``time'' variable $\eta$ is defined by the relation
\begin{equation}
    \partial\eta/\partial t\,=\, R^{-1}\beta^{1/2}\,,
\end{equation}
and where $t_{B}(r)$ is another free spatial function. The universe
age $T(r)$ corresponds to the time past since big-bang $R(r,\eta=0)=0$
at distance $r$ from the center and amounts to
\begin{align}
    T=t_{0}-t_{B}=\frac{1}{H_{\perp,0}}\left[\frac{1}{\Omega_{K0}}-\frac{\Omega_{m0}}{(\Omega_{K0})^{3/2}} {\rm arcsinh} \sqrt{\frac{\Omega_{K0}}{\Omega_{m0}}}\right],
\end{align}
 where $t_{0}$ is the present time. Of the four free spatial functions
that determine the solution, $t_{B}(r)$, $R_{0}$, $\Omega_{m0}(r)$ and $H_{\perp,0}$,
two can be fixed arbitrarily by a redefinition of $r$ and $t$. Henceforth, following most of the literature, we  choose $R_{0}=r$ and $t_{B}(r)=0$, i.e.,
adopt the same function that reproduces the FRW limit at the present epoch and synchronize the clocks at big-bang
time. The two remaining degrees of freedom can be expressed equivalently
by $\,\Omega_{m0}(r),H_{\perp0}\,$ or by $\,\alpha(r),\beta(r)\,$ or other
combinations. So we can write the relation
\begin{align}
    \alpha(r) & =R_{0}^{3}H_{\perp0}^{2}\Omega_{m0}\,\label{eq:alpha-Om0}\\
    \beta(r) & =R_{0}^{2}H_{\perp0}^{2}(1-\Omega_{m0})^{}\,,\label{eq:beta-Om0}
\end{align}
useful to convert models given in literature into one another. Fixing
the cosmic age $T(r)$ to be spatially homogeneous, one eliminates
yet another degree of freedom leaving only one free function. This also
ensures that there are no huge inhomogeneities in the past. For simplicity,
all the models we use below are chosen to have homogeneous cosmic
age but this choice plays no special role in what concerns our analysis.

\subsection{Current Constraints on Void Models}

\label{sec:ltb-constraints}

Void models have been studied quite intensively in the last few years
and several ideas have been put forward to constrain their properties.
We mentioned already the possibility of constraints due to spectral
distortions induced by scattered CMB light either from reionized regions~\cite{caldwell08}
or by the hot intracluster medium~\cite{garcia-bellido08b}. The
current data constrain the void size to be no larger than 1-2 Gigaparsecs,
although with a strong dependence on the central density and the profile.
In any case, voids this large are still a good fit of the supernovae
data (see e.g. the recent analyses of Refs.~\cite{alexander08,clifton08,flsc}).

Since in general we have two free functions, we need two independent
observables to reconstruct the void profile. The number density data are heavily subject
to evolution, selection and bias effects, so probably the most promising
method is to combine the estimation of angular or luminosity distances
(provided by supernovae or baryon acoustic oscillations) with a direct
measure of the expansion rate $H(z)$ given by longitudinal baryon
acoustic oscillations~\cite{clarkson08}, as suggested in~\cite{flsc}.


\subsection{Off-center Observers}

\label{sec:off-center}

Although most authors consider the observers to be at the center of symmetry of the LTB void for simplicity, there is no \emph{a priori} reason for that and one should consider the possibility of off-center observers. This has been done in~\cite{alnes06b,alnes07} and it was shown that supernovae and the size of the CMB dipole limit such a displacement to around $150$~\cite{alnes07} and $15$~Mpc~\cite{alnes06b} (in terms of the physical distance), respectively. Actually, as will be shown, a more accurate limit on the latter case is 26~Mpc, and a recent analysis showed that supernovae constraints may be a little looser~\cite{blomqvist09}. Nevertheless the current tightest constraints on void-induced anisotropies come from the CMB dipole.

However, in order to derive such a limit one has to assume that the
observer has no relative velocity relative to the surface of last
scattering. In other words, the CMB dipole would be completely due
to the off-center displacement. This is in direct contrast to the
standard FRW scenario, where the dipole is almost completely due to
our own peculiar velocity. If on the other hand the off-center observer
in LTB has a peculiar velocity, then the maximum off-center distance
can vary substantially. A good estimation of this limit can be done
following~\cite{alnes06b} through a simplified Newtonian picture,
which was numerically confirmed to give very good description.

The measured CMB dipole is $3.358\pm0.023\unit{mK}$~\cite{lineweaver96},
which when compared to the average CMB temperature of $2.725\unit{K}$
gives a temperature contrast $\Theta$ with an amplitude of 0.0012.
If the dipole is due only to the off-center displacement, one can
write (in the usual spherical harmonics decomposition)
\begin{equation}
    \Theta^{{\rm dipole}}=a_{10}Y_{10}\,,\label{eq:ThetaSH}
\end{equation}
from which one gets $a_{10}=2.5\cdot10^{-3}$ (note that in this
case $a_{11}$ and $a_{1-1}$ are both zero~\cite{alnes06b}). The
LTB off-center dipole seen by an observer at a physical distance $X_{{\rm obs}}$,
when compared to the homogeneous FRW case with a spatially constant
Hubble parameter $h^{{\rm out}}$, can be understood as an equivalent
``peculiar velocity'' of roughly
\begin{equation}
    \beta_{v}\equiv\frac{v_{p}}{c}=\frac{h^{{\rm in}}-h^{{\rm out}}}{3000\unit{Mpc}}X_{{\rm obs}}\label{eq:betav}
\end{equation}
with respect to the origin. In such a picture, the temperature anisotropies
measured by the observer are attributed to a Doppler shift of the
CMB photons due to his motion. In this picture it was shown~\cite{alnes06b}
that the dipole scales linearly, the quadrupole quadratically, and
the octopole cubically with the observers position. The expressions
for the dipole to the lowest order in $\beta_{v}$ is
\begin{align}
    a_{10} & =\sqrt{\frac{4\pi}{3}}\frac{h^{{\rm in}}-h^{{\rm out}}}{3000\unit{Mpc}}X_{{\rm obs}}\,.\label{eq:a10-newt}
\end{align}
From this approximation one gets that the maximum off-center physical
distance is $26\unit{Mpc}$.

This Newtonian picture is also very useful if we want to consider
both effects at the same time: an off-center distance and a (real)
peculiar velocity of the observer. Without a real peculiar velocity,
\eqref{eq:betav} gives $\beta_{v}=373\unit{km/s}$, which not surprisingly
is very close to the CMB inferred velocity between the Sun and the
CMB in a standard FRW metric. If one now considers a typical (real)
peculiar velocity of $500\unit{km/s}$ in the LTB case it is easy
to see from~\eqref{eq:betav} that if this velocity is in the direction
of the LTB center, one can have an effective $\beta_{v}^{{\rm eff}}=873\unit{km/s}$,
which pushes back the maximum off-center physical distance to a little
more that $60\unit{Mpc}$. Of course we have no reason to believe
that such an alignment should exist, but neither do we currently possess
any observations that could break such a degeneracy. In other words,
although not likely, an off-center distance of $60\unit{Mpc}$ cannot
currently be ruled out\footnote{In fact, a higher peculiar velocity of, say, $1000\unit{km/s}$ could
stretch this value to almost $100\unit{Mpc}$, after which other anisotropical
constraints such as the ones coming from supernovae are likely to
be more stringent.}. Nevertheless, in order to separate both void-induced and velocity-induced
effects, we will not push for such an aggressive off-center distance
and henceforth we will assume as the fiducial displacement a more
conservative value of $30\unit{Mpc}$ (although we will come back to this point in Section~\ref{sec:measure-cp}).

\section{Cosmic Parallax and Redshift Drift in LTB}

\label{sec:cp-and-zdot}

\subsection{Estimating the parallax for general anisotropy}

\begin{figure}[t!]
\includegraphics[width=7.5cm]{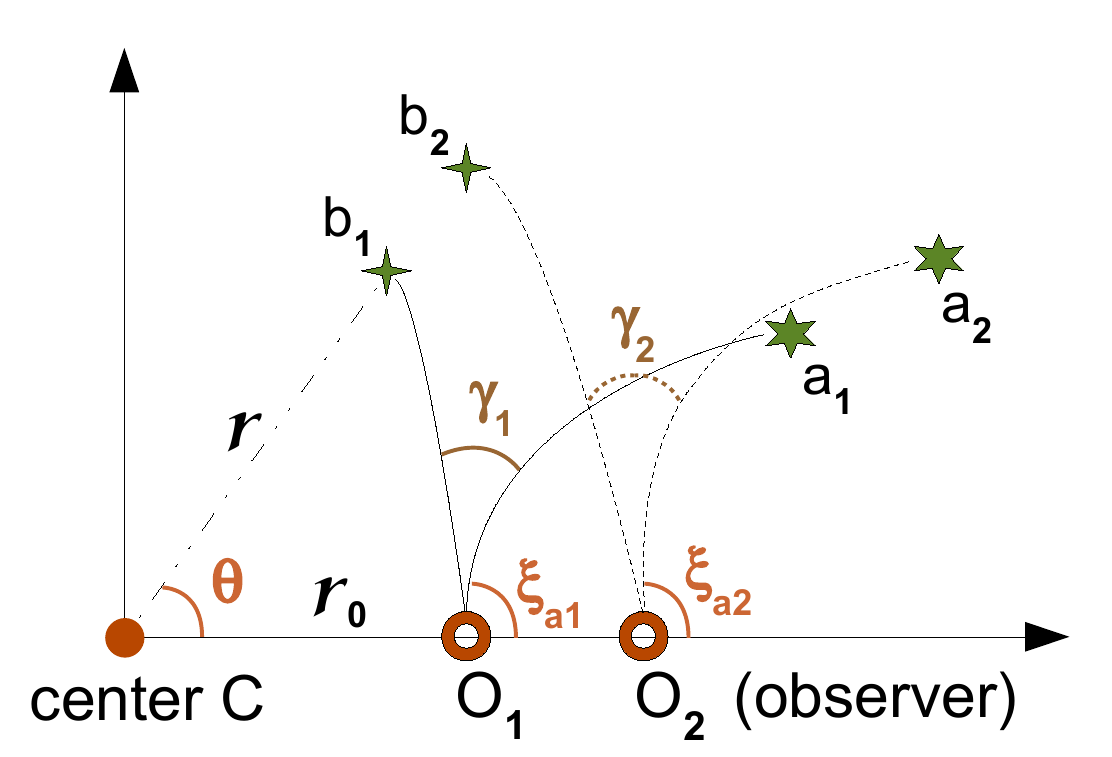}
\caption{Overview, notation and conventions of the cosmic parallax effect.
Note that for clarity purposes we assumed here that the points $C,O_{1},a_{1},b_{1}$
all lie on the same plane. By symmetry, points $O_{2},a_{2},b_{2}$
remain on this plane as well. Comoving coordinates $r$ and $r_{{\rm obs}}$
correspond to physical coordinates $X$ and $X_{{\rm obs}}$.}
\label{fig:overview}
\end{figure}

Figure~\ref{fig:overview} depicts the overall scheme describing
a possible time-variation of the angular position of a pair of sources
that expand anisotropically with respect to the observer. We label
the two sources $a$ and $b$, and the two observation times 1 and
2. In what follows, we will refer to ($t$, $r$, $\theta$, $\phi$)
as the comoving coordinates with origin on the center of a spherically
symmetric model. Peculiar velocities apart, the symmetry of such a
model forces objects to expand radially outwards, keeping $r$, $\theta$
and $\phi$ constant.

Let us assume now an expansion in a flat FRW space from a {}``center''
$C$ observed by an off-center observer $O$ at a distance $X_{{\rm obs}}$
from $C$. Since we are assuming FRW it is clear that any point in
space could be considered a {}``center'' of expansion: it is only
when we will consider a LTB universe that the center acquires an absolute
meaning. The relation between the observer line-of-sight angle $\xi$
and the coordinates of a source located at a radial distance $X$
and angle $\theta$ in the $C$-frame is
\begin{equation}
    \cos\xi=\frac{X\cos\theta-X_{{\rm obs}}}{(X^{2}+X_{{\rm obs}}^{2}-2\, X_{{\rm obs}}X\cos\theta)^{1/2}},
\end{equation}
 where all angles are measured with respect to the $CO$ axis and
all distances in this section are to be understood as physical distances.
Through most of this paper we shall assume for simplicity (and clarity)
that both sources share the same $\phi$ coordinate.

We consider first two sources at location $a_{1},b_{1}$ on the same
plane that includes the $CO$ axis with an angular separation $\gamma_{1}$
as seen from $O$, both at distance $X$ from $C$. After some time
$\Delta t$, the sources move to positions $a_{2},b_{2}$ and the
distances $X$ and $X_{{\rm obs}}$ will have increased by $\Delta_{t}X$
and $\Delta_{t}X_{{\rm obs}}$ respectively, so that the sources subtend
an angle $\gamma_{2}$. In a FRW universe, these increments are such
that they keep the overall separation $\gamma$ constant. However,
if for a moment we allow ourselves the liberty of assigning to the
scale factor $a(t)$ and the $H$ function a spatial dependence, a
time-variation of $\gamma$ is induced. The variation
\begin{equation}
    \Delta_{t}\gamma\;\equiv\;\gamma_{1}-\gamma_{2}
\end{equation}
is the cosmic parallax effect and can be easily estimated if we suppose that the Hubble law is just generalized to
\begin{equation}
    \Delta_{t}X=XH(t_{0},X)\Delta t\equiv XH_{X}\Delta t\,,
\end{equation}
where
\begin{equation}
    X(r)\equiv\int^{r}g_{rr}^{1/2}\dd r'=\int^{r}a(t_{0},r')\dd r'\,,\label{eq:phys-distance}
\end{equation}
generalizes the FRW relation $\, X_{{\rm FRW}}=a(t_{0})r$ in a metric
whose radial coefficient is $g_{rr}$.

For two arbitrary sources at distances much larger than $X_{{\rm obs}}$,
after straightforward geometry we arrive at
\begin{equation}
    \Delta_{t}\gamma\,=\,\Delta t(H_{{\rm obs}}-H_{X})X_{{\rm obs}}\left[\frac{\sin\theta_{a}}{X_{a}}-\frac{\sin\theta_{b}}{X_{b}}\right].\label{eq:deltagamma-full}
\end{equation}
For sources on similar shells, i.e., separated by a small $\Delta X\equiv X_{b}-X_{a}$
(not to be mistaken with the \emph{time} interval $\Delta_{t}X$),
we can write
\begin{equation}
\begin{aligned}
    \Delta_{t}\gamma\,\simeq\, s\,\Delta t\,(H_{{\rm obs}}-H_{X})\!\left[\sin\theta_{a}-\sin\theta_{b}\left(1-\frac{\Delta X}{X}\right)\right],
\end{aligned}
\label{eq:deltagamma-series}
\end{equation}
where we dropped the index {}``$a\,$'' on $X$, $H_{{\rm obs}}\equiv H(t_{0},r_{{\rm obs}})$
and we defined the parameter
\begin{equation}
    s\equiv\frac{X_{{\rm obs}}}{X}\ll1\,.\label{eq:s-definition}
\end{equation}
The above analytical estimates have been verified numerically, and
the angular dependence of the CP for sources at similar distances
has been verified to hold to very high precision.

The signal $\Delta_{t}\gamma$ in \eqref{eq:deltagamma-series} depends
on both source angles $\theta_{a,b}$. We can average over $\theta_{a,b}$
to obtain the average cosmic parallax for two arbitrary sources in
the sky (still assuming they lie on the same plane that contains $CO$).
If both sources are at the same redshift, then the average CP effect
is given by
\begin{align}
    \langle\Delta_{t}\gamma\rangle_{{\rm perp}} & \simeq\,\frac{s\,\Delta t\,(H_{{\rm obs}}-H_{X})}{4\pi^{2}}\int_{0}^{2\pi}\!\!\int_{0}^{2\pi}\!\!|\sin\theta_{a}-\sin\theta_{b}|\nonumber \\
     & \hspace{5.8cm}\dd\theta_{a}\dd\theta_{b}\,\label{eq:average-cp}\\
     & =\,\frac{8}{\pi^{2}}s\,\Delta t\,(H_{{\rm obs}}-H_{X})\,.
\end{align}

Note that at this order we can neglect the difference between the
observed angle $\xi$ and $\theta$. We can also convert the above
intervals $\Delta X$ into the redshift interval $\Delta z$ by using
the relation $\;r=\int\dd z/H(z)$. Using~\eqref{eq:phys-distance}
we can write $\;\Delta X=a(t_{0},X)\Delta z/H(z)\sim\Delta z/H(z)\,$
(we impose the normalization $\,a(t_{0},X_{{\rm obs}})=1$), where $\,H(z)\equiv H(t(z),X)$.
One should note that in a non-FRW metric, one has $\,s\neq r_{0}/r$.

In a FRW metric, $H$ does not depend on $r$ and the parallax vanishes.
On the other hand, any deviation from FRW entails such spatial dependence
and the emergence of cosmic parallax, except possibly for special
observers (such as the center of LTB). A constraint on $\Delta_{t}\gamma$
is therefore a constraint on cosmic anisotropy.

Rigorously, the use of the above equations is inconsistent outside
a flat FRW scenario; one actually needs to perform a full integration
of light-ray geodesics in the new metric. Nevertheless, we shall assume
for a moment that for an order of magnitude estimate we can simply
replace $H$ with its space-dependent counterpart given by LTB models.
In order for an alternative LTB cosmology to have any substantial
effect (e.g., explaining the SNIa Hubble diagram) it is reasonable
to assume a difference between the local $H_{{\rm obs}}$ and the
distant $H_{X}$ of order $H_{{\rm obs}}$~\cite{alnes06b}. More
precisely, putting $\, H_{obs}-H_{X}=H_{obs}\Delta h\,$ then using
\eqref{eq:average-cp} one has that the average $\Delta_{t}\gamma$
is of order
\begin{equation}
    \langle\Delta_{t}\gamma\rangle\Big|_{{\rm perp}}\sim\,20\, s\Delta h\,\unit{\mu as/year}\label{eq:perp-cp-estimate}
\end{equation}
 for two sources at the same redshift. Similarly, for source pairs
at same position $\theta$ but different (yet similar) redshifts one
has (using~\eqref{eq:deltagamma-series})
\begin{equation}
\begin{aligned}
    \Delta_{t}\gamma\Big|_{{\rm rad}} & \sim\, s\,\sin\theta\Delta h\Delta t\,\Delta z/X\;\mu\mbox{as/year}\\
     & \sim\,20\, s\sin{\theta}\,\Delta h\frac{\Delta z}{z}\;\mu\mbox{as/year}\,,
\end{aligned}
\end{equation}
 where it was assumed that $X\sim zH(z)^{-1}$. The average radial
CP for sources between 10 and 200 times $X_{{\rm obs}}$ can be obtained
numerically to be
\begin{align}
    \langle\Delta_{t}\gamma\rangle_{{\rm rad}} & \simeq\,\frac{\,\Delta t\,(H_{{\rm obs}}-H_{X})\sin\theta}{190^{2}}\int_{10}^{200}\!\!\int_{10}^{200}\!\left|\frac{1}{s_{a}}-\frac{1}{s_{b}}\right|\nonumber \\
     & \hspace{4.5cm}\dd (1/s_{a})\,\dd (1/s_{b})\,\label{eq:average-rad-cp}\\
     & =\,0.014\,\sin\theta\Delta t\,(H_{{\rm obs}}-H_{X})\,.\label{eq:deltagamma}
\end{align}
Therefore, one can estimate for the radial signal
\begin{equation}
    \langle\Delta_{t}\gamma\rangle\Big|_{{\rm rad}}\sim\,0.3\sin\theta\Delta h\,\unit{\mu as/year}\,,\label{eq:rad-cp-estimate}
\end{equation}
 which is very similar to its same-shell counterpart~\eqref{eq:perp-cp-estimate},
except for the $\sin\theta$ modulation.

Let us finally consider the main expected source of noise, the intrinsic
peculiar velocities of the sources. The variation in angular separation
for sources at angular diameter distance $D_{A}$ (measured by the
observer) and peculiar velocity $v_{{\rm pec}}$ can be estimated
as
\begin{equation}
    \Delta_{t}\gamma_{{\rm pec}}=\left(\frac{v_{{\rm pec}}}{500\,\frac{\mbox{km}}{\mbox{s}}}\right)\left(\frac{D_{A}}{1\,\mbox{Gpc}}\right)^{-1}\!\left(\frac{\Delta t}{10\,\mbox{years}}\right)\mu\mbox{as}.\label{eq:pecvel}
\end{equation}
This velocity field noise is therefore typically smaller than the
experimental uncertainty (especially for large distances) and again
will be averaged out for many sources. Notice that the observer's
own peculiar velocity produces a systematic offset sinusoidal signal
$\Delta_{t}\gamma_{{\rm pec},O}$ of the same amplitude as $\Delta_{t}\gamma_{{\rm pec}}$
that has to be subtracted from the observations: we discuss this further
below. The above relation was further investigated in~\cite{ding09}, where it was proposed to estimate $D_{A}$ via observations
of $\Delta_{t}\gamma_{{\rm pec}}$ due not to voids but by our motion with respect to the CMB.

\subsection{Geodesic Equations}

\label{sec:geodesic}

As suggestive as the above estimates be, they need confirmation from
an exact treatment where the full relativistic propagation of light
rays is taken into account. We will thus consider in what follows
three particular LTB models capable of fitting the observed SNIa Hubble
diagram and the CMB first peak position and compatible with the COBE
results of the CMB dipole anisotropy, as long as the observer is within
around 30 Mpc from the center~\cite{alnes06b}. Moreover, all three
models have void sizes which, although huge by any means, are {}``small''
enough ($z \sim 0.3 - 0.4$) not to be ruled out due to distortions of the
CMB blackbody radiation spectrum~\cite{caldwell08}.

Due to the axial symmetry and the fact that photons follow a path
which preserves the 4-velocity identity $u^{\alpha}u_{\alpha}=0$,
the four second-order geodesic equations for $(t,r,\theta,\phi)$,
\begin{equation}
    \frac{\dd^{2}x^{a}}{\dd\lambda^{2}}+\Gamma^{a}{}_{bc}\frac{\dd x^{b}}{\dd\lambda}\frac{\dd x^{c}}{\dd\lambda}=0\,,\label{eq:general-geodesic}
\end{equation}
 can be written as five first-order ones. Here $\lambda$ is the arbitrary
affine parameter of the geodesics. We will choose as variables the
center-based coordinates $t$, $r$, $\theta$, $p\equiv\dd r/\dd\lambda$
and the redshift $z$. We shall refer also to the conserved angular
momentum
\begin{equation}
    J\equiv R^{2}\frac{\dd\theta}{\dd\lambda}=\textrm{\emph{const}}=J_{0}\,,\label{eq:J}
\end{equation}
 which is a direct consequence of the $a\rightarrow\theta$ equation in~\eqref{eq:general-geodesic}.
For a particular source, the angle $\xi$ is the coordinate equivalent
to $\theta$ for the observer, and in particular $\xi_{0}$ is the
coordinate $\xi$ of a photon that arrives at the observer at the
time of observation $t_{0}$. Obviously this coincides with the measured
position in the sky of such a source at $t_{0}$. In terms of these
variables, and defining $\lambda$ such that $u(\lambda)<0$, the
autonomous system governing the geodesics is written as (see~\cite{alnes06b})
\begin{equation}
\begin{aligned}
    \frac{\dd t}{\dd\lambda}\;=\; & -\sqrt{\frac{(R')^{2}}{1+\beta}\, p^{2}+\frac{J^{2}}{R^{2}}}\,,\\
    \frac{{\textrm{d}}r}{{\textrm{d}}\lambda}\;=\; & p\,,\quad\quad\frac{\dd\theta}{\dd\lambda}\;=\;\frac{J}{R^{2}}\,,\hspace{-3cm}\\
    \frac{\dd z}{\dd\lambda}\;=\; & \frac{(1+z)}{\sqrt{\frac{(R')^{2}}{1+\beta}\, p^{2}+\frac{J^{2}}{R^{2}}}}\left[\frac{R'\dot{R}'}{1+\beta}\, p^{2}+\frac{\dot{R}}{R^{3}}J^{2}\right],\\
    \frac{\dd p}{\dd\lambda}\;=\; & 2\dot{R}'\, p\,\sqrt{\frac{p^{2}}{1+\beta}\,+\frac{J^{2}}{R^{2}\, R'^{2}}}\,+\,\frac{1+\beta}{R^{3}R'}J^{2}\,+\\
     & +\left[\frac{\beta'}{2+2\beta}-\frac{R''}{R'}\right]p^{2}\,.
\end{aligned}
\label{eq:geodesics}
\end{equation}

The angle $\xi$ along a geodesic is given by~\cite{alnes06b}:
\begin{equation}
    \cos\xi=-\frac{R'(t,r)\, p}{u\,\sqrt{1+\beta(r)}}\,,
\end{equation}
 from which we obtain, exploiting the remaining freedom in the definition
of $\lambda$, the relations~\cite{alnes06b}
\begin{align}
    p_{0} & \;=\;-\frac{\sqrt{1+\beta(r_{{\rm obs}})}}{R'(t_{0},r_{{\rm obs}})}\cos(\xi_{0})\,,\label{eq:p0-and-J0}\\
    J_{0} & \;=\; J\;=\; R(t_{0},r_{{\rm obs}})\,\sin(\xi_{0})\,.
\end{align}
 Therefore, our autonomous system is completely defined by the initial
conditions $t_{0}$, $r_{{\rm obs}}$, $\theta_{0}=0$, $z_{0}=0$
and $\xi_{0}$. The first two define the instant of measurement and
the offset between observer and center, while $\xi_{0}$ stands for
the direction of incidence of the photons.

An algorithm for predicting the variation of an arbitrary angular
separation can be written as follows:
\begin{enumerate}
\item Denote with $(z_{a1},z_{b1},\xi_{a1},\xi_{b1})$ the observed coordinate
of a pair of sources at a given time $t_{0}$ and observer position
$r_{{\rm obs}}$; \vspace{-0.1cm}

\item Solve numerically the autonomous system with initial conditions $(t_{0},r_{{\rm obs}},\theta_{0}=0,z_{0}=0,\xi_{0}=\xi_{a1})$
and find out the values of $\lambda_{a}^{\ast}$ such that $z(\lambda_{a}^{\ast})=z_{a1}\,$;
\vspace{-0.1cm}

\item Take note of the values $r_{a1}(\lambda_{a}^{\ast})\,$ and $\theta_{a1}(\lambda_{a}^{\ast})\,$
(since the sources are assumed comoving with no peculiar velocities,
these values are constant in time); \vspace{-0.1cm}

\item Define $\lambda_{a}^{\dagger}$ as the parameter value for which $r_{a2}(\lambda_{a}^{\dagger})=r_{a1}(\lambda_{a}^{\ast})$,
where $r_{a2}$ is the geodesic solution for a photon arriving 
 $\Delta t$ later with an incident angle $\xi_{a2}$, and vary $\xi_{a2}$
until $\theta_{a2}(\lambda_{a}^{\dagger})=\theta_{a1}(\lambda_{a}^{\ast})\,$;
\vspace{-0.1cm}

\item Repeat the above steps for source $b$, and compute the difference
$\Delta_{t}\gamma\equiv\gamma_{2}-\gamma_{1}=(\xi_{a2}-\xi_{b2})-(\xi_{a1}-\xi_{b1})$.
\end{enumerate}
\vspace{-0.1cm}

The above algorithm gives as a byproduct another interesting observable,
the Sandage redshift drift~\cite{sandage62,loeb98} (see Section~\ref{sec:zdot}). It is important to realize
that the redshift drift is inherently coupled to the CP, that is,
in principle one cannot calculate one effect without taking the other
into account. A general prescription on how to obtain $\dot{z}$ for an observer
at the center of a LTB model was obtained in~\cite{uzan08}. As we will show in Section~\ref{sec:numerical-results},
in the limit of small $\;s\;$ our numerical results reveal that $\Delta_t z$
for off-center observers show small angular dependence, and therefore
to good approximation one can neglect the CP effect when calculating
the redshift drift.

A remark on the above procedure is in order before we continue. Due
to the intrinsically smallness of both the cosmic parallax and Sandage
effects (in the course of a decade), a carefully constructed numerical
code is needed to correctly compute either. To give an idea of the
amount of precision required, consider the following: if one naively
calculates $\Delta_{t}\gamma$ for a $\Delta t$ of 10~years, one
needs to evaluate $\xi_{a1}$ and $\xi_{a2}$ with at least 13 digits
of precision (as the CP is of the order of $0.2\,\mu\mbox{as}\sim10^{-12}$~rad).
Although it is possible to alleviate this by exploiting the linearity
of $\Delta_{t}\gamma$ in $\Delta t$ and scaling up the system, it
still remains a numerically challenging problem, as was independently
found out in~\cite{fontanini09}. In Appendix~\ref{sec:numerical-nuances}
we explore this issue in more detail and describe how we were able
to circumvent it in both the present and original paper~\cite{quercellini09a}.

\begin{figure}[t]
 \includegraphics[width=7.5cm]{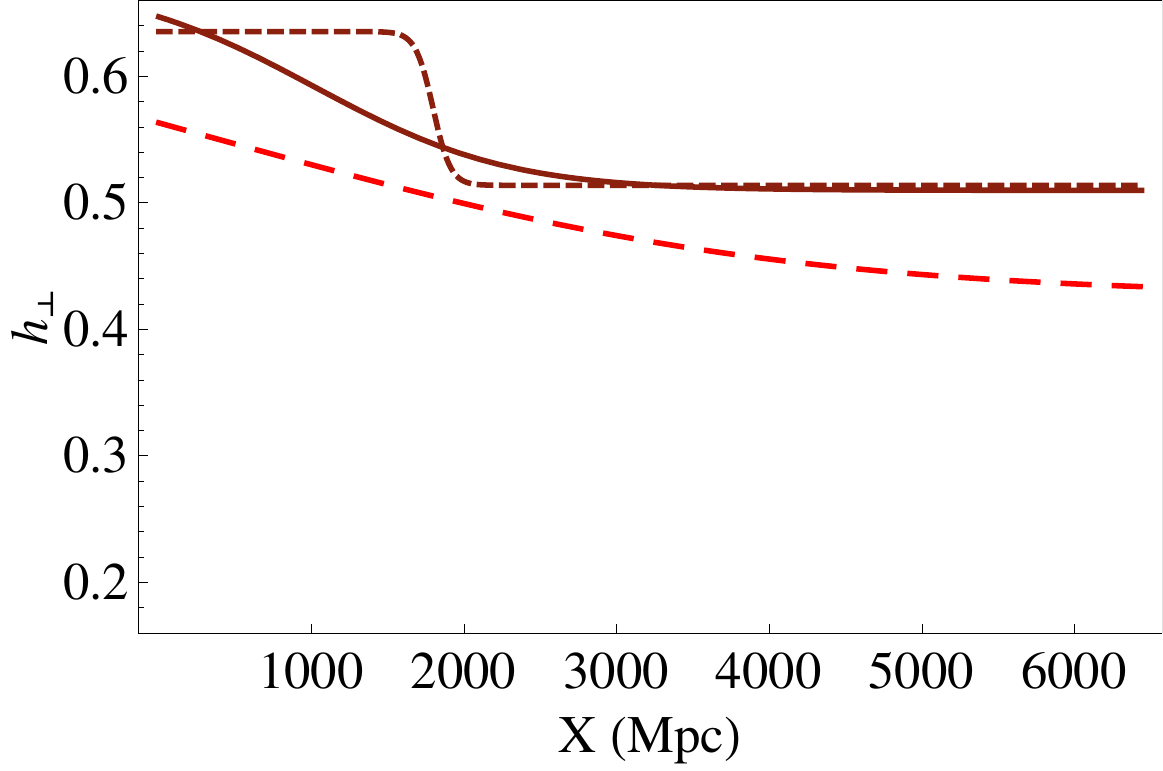}
 \includegraphics[width=7.5cm]{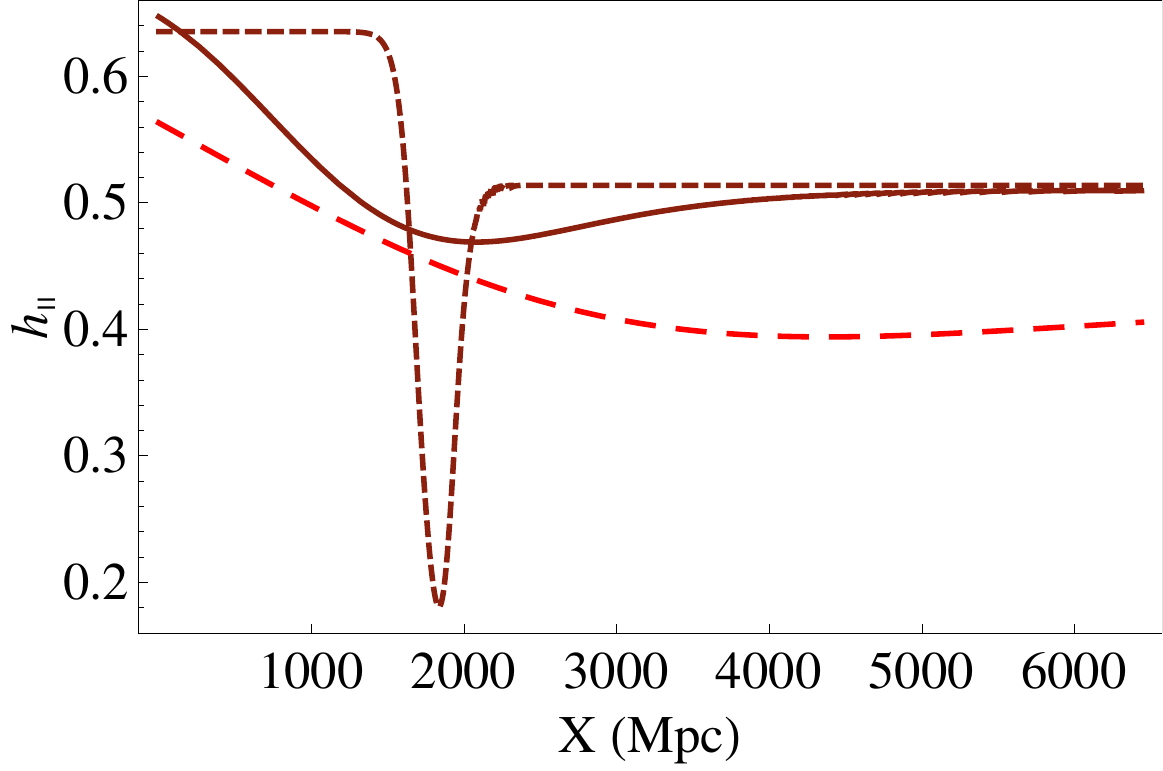}
\caption{$H_{||}$ and $H_{\perp}$ for Model I (solid), Model II (dashed curve)
and the cGBH model (red, long-dashed) in units of 100\unit{km /
(s Mpc)}, as a function of the physical distance $X$. Note that
both Hubble parameters differ only around the void-transition region.}
\label{fig:ltb-Hs}
\end{figure}

\subsection{Specific Models}

\label{sec:models}

The models of Refs.~\cite{alnes06b,alnes07} are characterized by a smooth
transition between an inner void and an outer region with higher matter
density and described by the functions:
\begin{align}
    \alpha(r) & =\left(H_{\perp,0}^{{\rm out}}\right)^{2}r^{3}\left[1-\frac{\Delta\alpha}{2}\left(1-\tanh\frac{r-r_{{\rm vo}}}{2\Delta r}\right)\right]\!,\label{eq:alpha}\\
    \beta(r) & =\left(H_{\perp,0}^{{\rm out}}\right)^{2}r^{2}\,\frac{\Delta\alpha}{2}\left(1-\tanh\frac{r-r_{{\rm vo}}}{2\Delta r}\right)\!,\label{eq:beta}
\end{align}
where $\Delta\alpha$, $r_{{\rm vo}}$ and $\Delta r$ are three
free parameters and $H_{\perp,0}^{{\rm out}}$ is the Hubble constant
at the outer region, set at 51~km~s$^{-1}$~$\mbox{Mpc}^{-1}$.

We will dub the two models I and II, and define them by the sets $\{\Delta\alpha=0.9,\, r_{{\rm vo}}=1.46\mbox{ Gpc},\Delta r=0.4\, r_{{\rm vo}}\}$ and $\{\Delta\alpha=0.78,r_{{\rm vo}}=1.83$~Gpc$,\Delta r=0.03\, r_{{\rm vo}}\}$,
respectively. These values of $r_{{\rm vo}}$ correspond, in physical
distances, to void sizes of 1.34 and 1.68 Gpc, respectively. We will
also consider the so-called {}``constrained'' model proposed in~\cite{garcia-bellido08a}
which we will henceforth refer to as the ``cGBH'' model. For this
model, we choose the parameters that maximize the likelihood as obtained
in~\cite{garcia-bellido08a}, which can be written in terms of $\alpha$
and $\beta$ using~\eqref{eq:alpha-Om0} and~\eqref{eq:beta-Om0}.
The main difference between the three models is that Model~II features
a much sharper transition from the void and that the cGBH model is almost
twice as large. Nonetheless, neither transition width nor void size
are expected to be important factors in cosmic parallax since in any
case most quasars are outside the void and the most relevant quantity
is the difference between the inner and outer values of $H$. In all
three cases we set the off-center (physical) distance to 30~Mpc,
which is the upper limit allowed by CMB dipole distortions (see Section~\ref{sec:off-center}),
and this corresponds to $s\simeq9\;10^{-3}$ for a source at $z=1$.

\begin{figure}[t]
     \includegraphics[width=7.5cm]{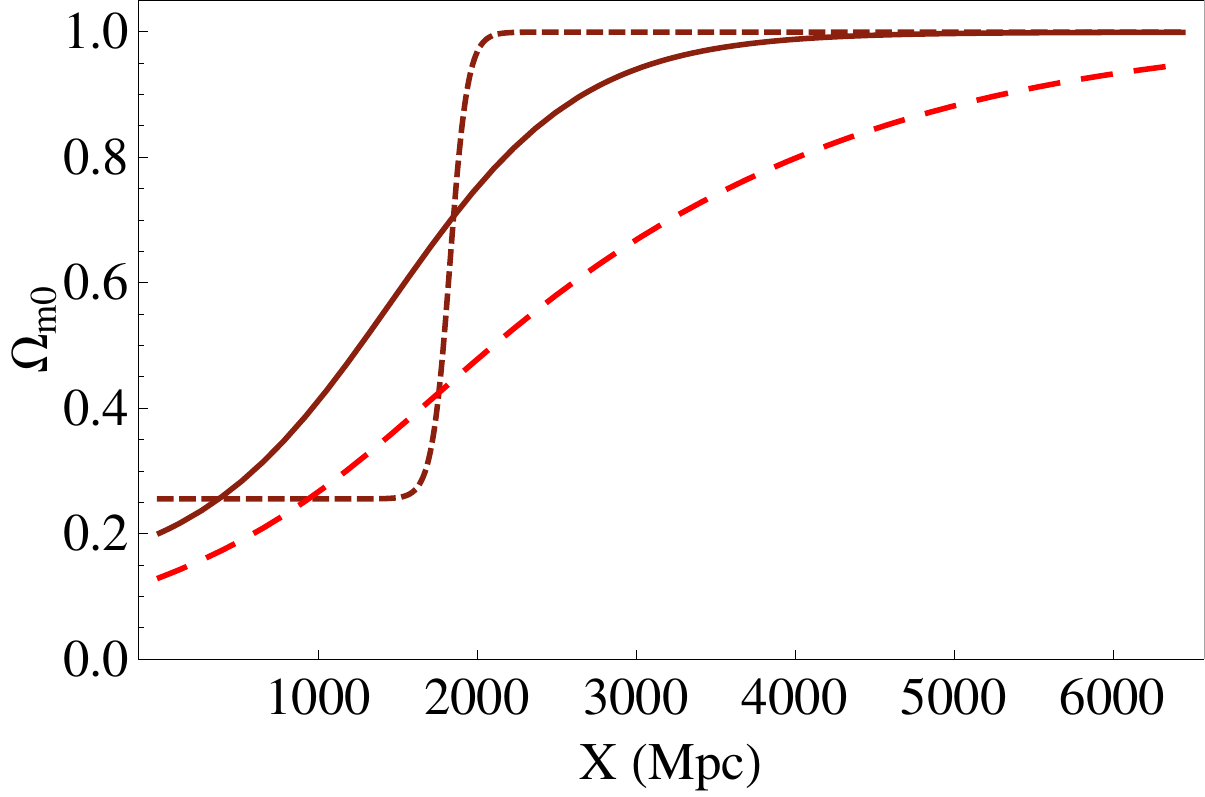}
    \caption{$\Omega_{m0}$ for Model I (solid), Model II (dashed curve) and the
    cGBH model (red, long-dashed) as a function of the physical distance
    $X$. Note that the definition for $\Omega_{m0}$ we use differ from
    the one in~\cite{alnes06a,alnes06b,alnes07} and we do not get their
    characteristic over-density bump (or shell) surrounding the void.}
    \label{fig:ltb-omega-m}
\end{figure}

Figure~\ref{fig:ltb-Hs} depicts the behavior of $H_{\perp,0}$ and
$H_{||,0}$ as a function of the comoving distance from the center
of the void. Note that overall both functions are similar, specially
outside the void (where they quickly approach $0.5$). The main discrepancy
is seen in $H_{||,0}$ for Model II around the (sharp) transition
region of the void. Similarly, figure~\ref{fig:ltb-omega-m} illustrates
the void by depicting $\Omega_{m0}$ from the inside to the outside
region, where it evaluates to unity.

In order to make better use of the FRW-like estimates in an LTB universe,
one must first understand which $H$, parallel or transverse, corresponds
to $H_{{\rm obs}}$ and $H_{X}$ in~\eqref{eq:deltagamma-full}.
From~\eqref{eq:phys-distance} we get
\begin{equation}
    X_{{\rm LTB}}=\int^{r}\frac{R'(t_{0},r')}{\sqrt{1+\beta(r')}}\,\dd r'\,
\end{equation}
 and
\begin{equation}
\begin{aligned}
    & \frac{\partial X}{\partial t}\;=\;\int^{r}\frac{\dot{R}'(t_{0},r')}{\sqrt{1+\beta(r')}}\,\dd r'\\
    & \quad\stackrel{\mbox{{\footnotesize eq. }}\eqref{eq:def-Hpar}}{=}\;\int^{r}H_{||}(t_{0},r')\frac{R'(t_{0},r')}{\sqrt{1+\beta(r')}}\,\dd r'\,.
\end{aligned}
\end{equation}
 Therefore we write $\Delta X=(\partial X/\partial t)\,\Delta t\,+\,{\cal O}(\Delta t^{2})$
and thus, defining $\bar{H}$ such that to first order $\Delta X\equiv X\bar{H}\Delta t$,
we get \begin{equation}
\begin{aligned}
    \bar{H}\; & \;=\frac{1}{X}\int_{0}^{r}H_{||}(t_{0},r')\frac{R'(t_{0},r')}{\sqrt{1+\beta(r')}}\,\dd r'\\
    & \;=\frac{1}{\int_{0}^{r}\frac{R'(t_{0},r')}{\sqrt{1+\beta(r')}}\,\dd r'}\int_{0}^{r}H_{||}(t_{0},r')\frac{R'(t_{0},r')}{\sqrt{1+\beta(r')}}\,\dd r'\,.
\end{aligned}
\label{eq:H-bar}
\end{equation}
 In a step-like LTB void model ($\Delta r\rightarrow0$) the quantity
$H_{X}$ in~\eqref{eq:deltagamma} is given by
\begin{equation}
    H_{||}(t_{0},r)=H_{||}^{{\rm in}}+(H_{||}^{{\rm out}}-H_{||}^{{\rm in}})\Theta(r-r_{{\rm vo}})\,,
\end{equation}
 where $\Theta$ is the Heaviside (or {}``step'') function. Substituting
in~\eqref{eq:H-bar}, we finally arrive at the sought after result
$\, H_{{\rm obs}}=H_{||,0}^{{\rm in}}\,$ and
\begin{equation}
    \bar{H}=H_{||}^{{\rm in}}\,\frac{X_{{\rm vo}}}{X}+H_{||,0}^{{\rm out}}\,\left(1-\frac{X_{{\rm vo}}}{X}\right)\simeq H_{||,0}^{{\rm out}}\,.\label{eq:H-bar-result}
\end{equation}
 This shows that in general, the values of $H_{{\rm obs}}$ and $H_{X}$
in~\eqref{eq:deltagamma} are obtained by a combination of both $\, H_{||}\,$
and $\, H_{\perp}$. On all three models here considered, however, these
quantities differ by less than 30\%. Since our main goal is to use~\eqref{eq:deltagamma}
as an estimate of the true (numerical) effect, either one could be
used. Nevertheless, in order to be as accurate as possible we shall
(motivated by~\eqref{eq:H-bar-result}), substitute $H_{{\rm obs}}$
and $H_{X}$ by their LTB $\, H_{||}\,$ counterparts.

\subsection{The Sandage Redshift Drift}

\label{sec:zdot}

It has been known for a long time~\cite{sandage62} that for any
expanding cosmology the redshift $z$ of a given source is not a constant
in time. In a decelerating universe all redshifts decrease in time.
In models predicting a recent (since $z\sim1$) acceleration, like
the \lcdm\ model, sources with redshifts $\,z\lesssim 2\,$ actually have
positive $\,\dd z/\dd t$. In effect, observation of $\,\dd z/\dd t\,$
gives one a direct measurement of the expansion of the universe, and
is at least in principle one of the few direct ways of measuring directly
$H(z)$ (along with e.g. longitudinal acoustic oscillations).
The prospect of doing so was revisited in~\cite{loeb98}.

If one assumes a FRW metric, the observed redshift of a given source,
which emitted its light at a time $t_{s}$, is, today ($t_{0}$),
\begin{equation}
    z_{s}(t_{0})=\frac{a(t_{0})}{a(t_{s})}-1,
\end{equation}
 and it becomes, after a time interval $\Delta t_{0}$ ($\Delta t_{s}$
for the source)
\begin{equation}
    z_{s}(t_{0}+\Delta t_{0})=\frac{a(t_{0}+\Delta t_{0})}{a(t_{s}+\Delta t_{s})}-1.
\end{equation}
The observed redshift variation of the source is, then,
\begin{equation}
    \Delta_{t}z_{s}=\frac{a(t_{0}+\Delta t_{0})}{a(t_{s}+\Delta t_{s})}-\frac{a(t_{0})}{a(t_{s})},
\end{equation}
which can be re-expressed, after an expansion at first order in $\Delta t/t$,
as:
\begin{equation}
    \Delta_{t}z_{s}=\Delta t_{0}\left(\frac{\dot{a}(t_{0})-\dot{a}(t_{s})}{a(t_{s})}\right)+\mathcal{O}\left(\frac{\Delta t_{0}}{t_{0}}\right)^{2}.
\end{equation}
We can rewrite the last expression in terms of the Hubble parameter
$H(z)=\dot{a}(z)/a(z)$:
\begin{equation}
    \Delta_{t}z_{s}=H_{0}\Delta t_{0}\left(1+z_{s}-\frac{H(z_{s})}{H_{0}}\right).\label{deltaz}
\end{equation}
It will prove useful in Section~\ref{sec:measure-zdot}, where we
estimate achievable observational precision, to relate the redshift
variation to an apparent velocity shift of the source, $\Delta v=c\Delta_{t}z_{s}/(1+z_{s})$.

This redshift drift, or $\dot{z}$, or ``Sandage effect'', has
been investigated for a variety of dark energy models currently pursued
in the literature~\cite{chm,balbi07,cristiani07,liske08}, and it
is interesting to note that most of them predict a very similar redshift
profile for the effect, all very close to the one generated by the
$\Lambda$CDM model. In $\Lambda$CDM, the redshift drift is positive
in the region $\,0<z<2.4\,$ but becomes negative for higher redshift
(see Figure~\ref{fig:zdot-per-year}). On the other hand, a dark-energy
mimicking giant void produces a very distinct $z$ dependance of this
drift, and in fact one has, as we will show below, that $\dd z/\dd t$
is always negative.\footnote{It has recently come to our attention that this property and its potential as discriminator between LTB voids and \lcdm\ was first pointed out in~\cite{yoo08}.}

\begin{figure}[t]
    \includegraphics[width=7.8cm]{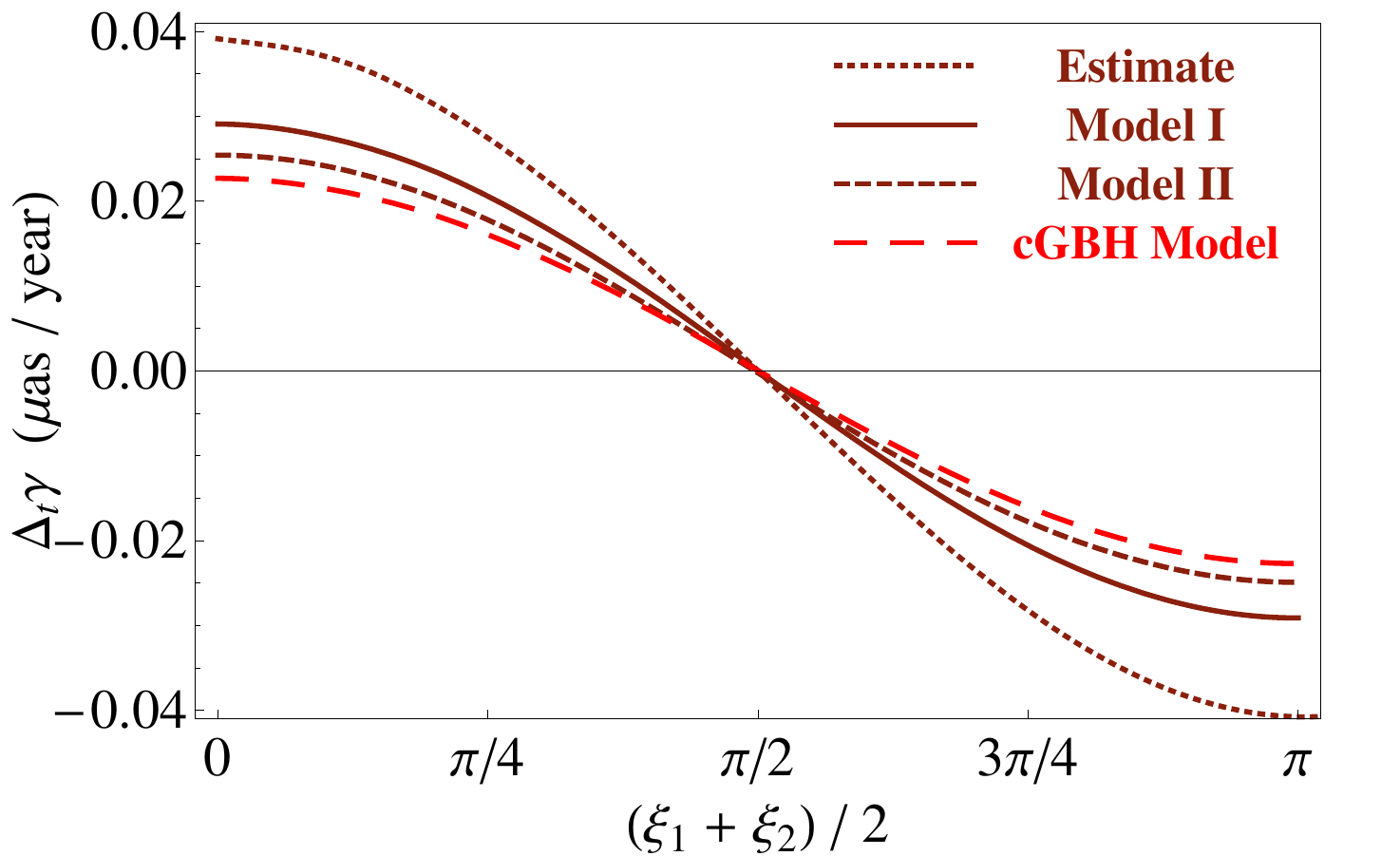}
    \caption{$\Delta_{t}\gamma$ for two sources at the same shell, at $z=1$,
    for Model I (full lines), Model II (dashed), the cGBH model (red,
    long-dashed lines) and the FRW-like estimate (dotted). The lines correspond
    to a separation of $90^{\circ}$ in the sky between the sources. The
    off-center distance is assumed to be $30\unit{Mpc}$.}
    \label{fig:dgammath}
\end{figure}

\subsection{Numerical Results}

\label{sec:numerical-results}

In Figure~\ref{fig:dgammath} we plot $\Delta_{t}\gamma$ for two
sources at $z=1$, for models I and II as well as for the FRW-like
estimate. One can see that the results do not depend sensitively on
the details of the shell transition and that in both cases the FRW-like
estimate gives a reasonable idea of the true LTB behavior. We conclude
that~\eqref{eq:deltagamma-full} is indeed a valid approximation.

\begin{figure}[t]
\includegraphics[width=8cm]{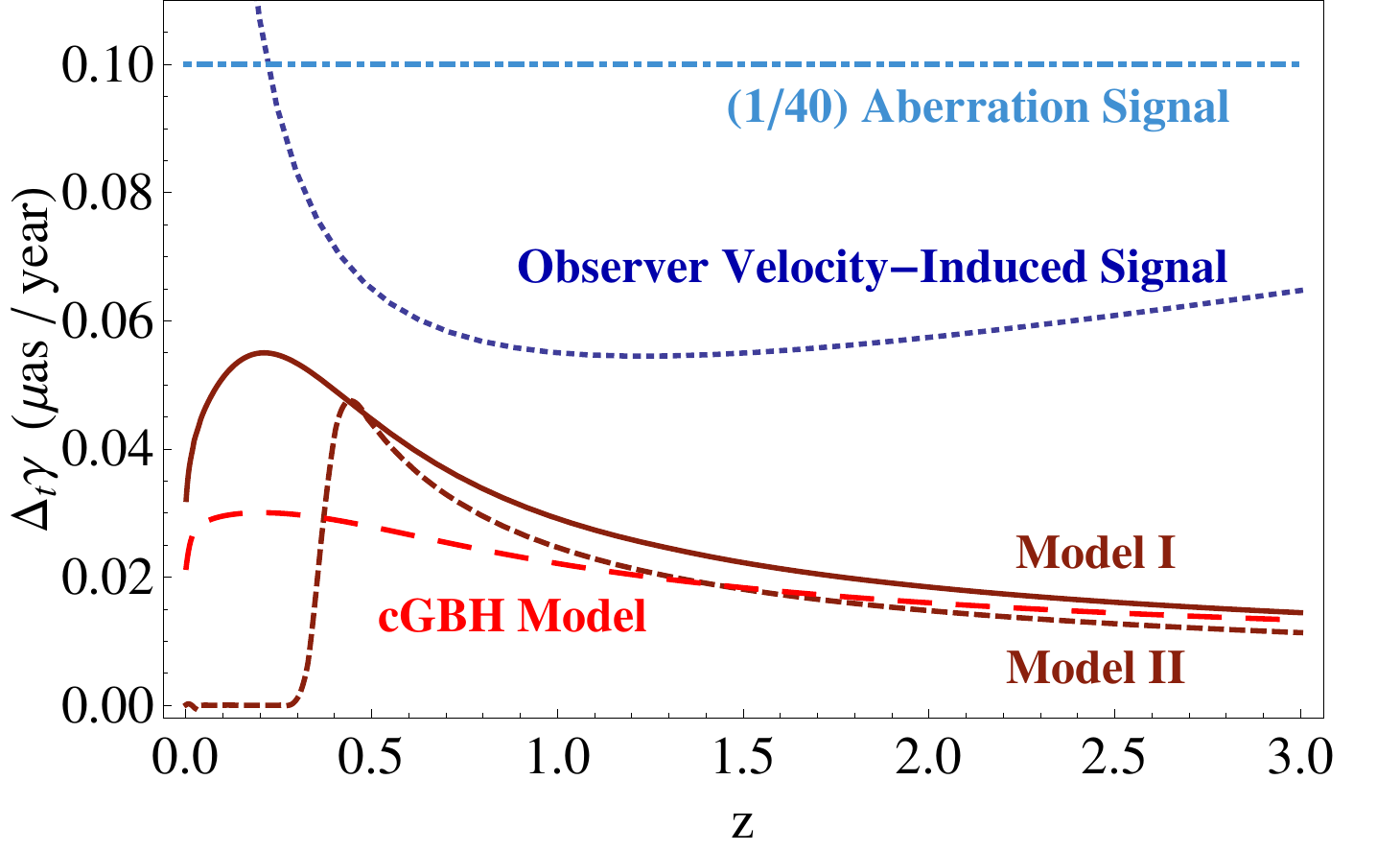}
\caption{$\Delta_{t}\gamma$ for two sources at the same shell but separated
by $90^{\circ}$ as a function of redshift assuming a $30\unit{Mpc}$
off-center distance. The dark, brown lines correspond to the cosmic
parallax in Models I (full lines) and II (dashed); the red long-dashed
lines to the cGBH model; the light, blue dotted lines represent
$1/40$ of the aberration-induced signal (see text), which does not depend
on redshift; the dark dotted lines stand for the parallax induced
by our own peculiar velocity (assumed to be 400 km/s).
Since all effects are dipolar, the curves plotted here are proportional
to the amplitude of such dipoles. The actual amount of noise depend
on the angle between the center of the void and the directions of
acceleration and peculiar velocity of the measuring instrument. Notice
that as expected, in Model II the CP is zero inside the
void.}
\label{fig:dgammath-vs-z}
\end{figure}

Figure~\ref{fig:dgammath-vs-z} depicts the redshift dependance of
the cosmic parallax effect for two sources at the same shell (i.e.,
same redshift) but separated in the sky by $90^{\circ}$ (which is
the average separation between two sources in an all-sky survey):
one source is located at $\xi=-45^{\circ}$, the other at $\xi=+45^{\circ}$.
Also plotted are the two major sources of systematic noise, which
will be discussed in Section~\ref{sec:measure-cp}: our own peculiar
velocity and the change in the aberration of the sky due to the acceleration
of the observer. As will be shown, all the effects we are considering
are dipolar and the lines in Figure~\ref{fig:dgammath-vs-z} are
proportional to the amplitudes of such dipoles. Note that both systematics
have different $z$-dependance than the CP produce in void models,
and in principle all three effects can be separated.

\begin{figure}[t]
 \includegraphics[width=8cm]{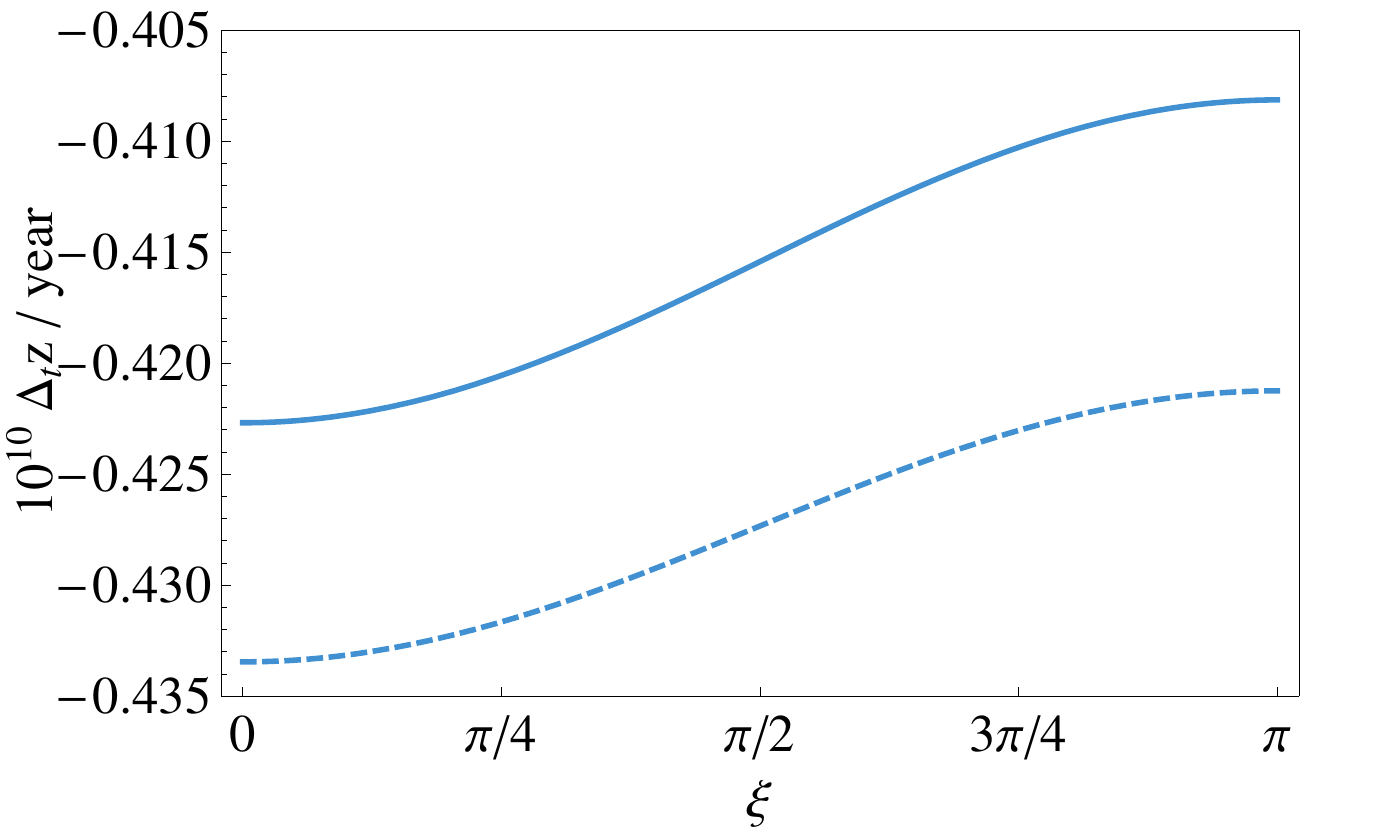}
\caption{The Sandage effect for a source at $\, z=1\,$ for an observer $30\unit{Mpc}$
away from the center as a function of the angle $\xi$ for both Models
I (full) and II (dashed lines). Note that the fractional fluctuation
of the redshift drift in the sky is less than 5\%, and one can therefore
assume isotropy to good precision when computing this effect on void
models.}
\label{fig:zdot-vs-angle}
\end{figure}

As mentioned before, in principle the Sandage effect and cosmic
parallax are two coupled effects and rigourously any calculation of
one effect must take into account the other. Nevertheless, in practice
the coupling is a weak one, and to compute the redshift drift one
can always assume to good precision that the observer is in the center
of the void. Figure~\ref{fig:zdot-vs-angle} illustrates this fact
by depicting the Sandage effect for a source at $\, z=1\,$ as a function
of the angle $\xi$ for both Models I and II, for an observer $30\unit{Mpc}$
away from the center. As can be seen, the fractional fluctuation of
the redshift drift in the sky is less than 5\%.

\begin{figure}[t]
\includegraphics[width=7.7cm]{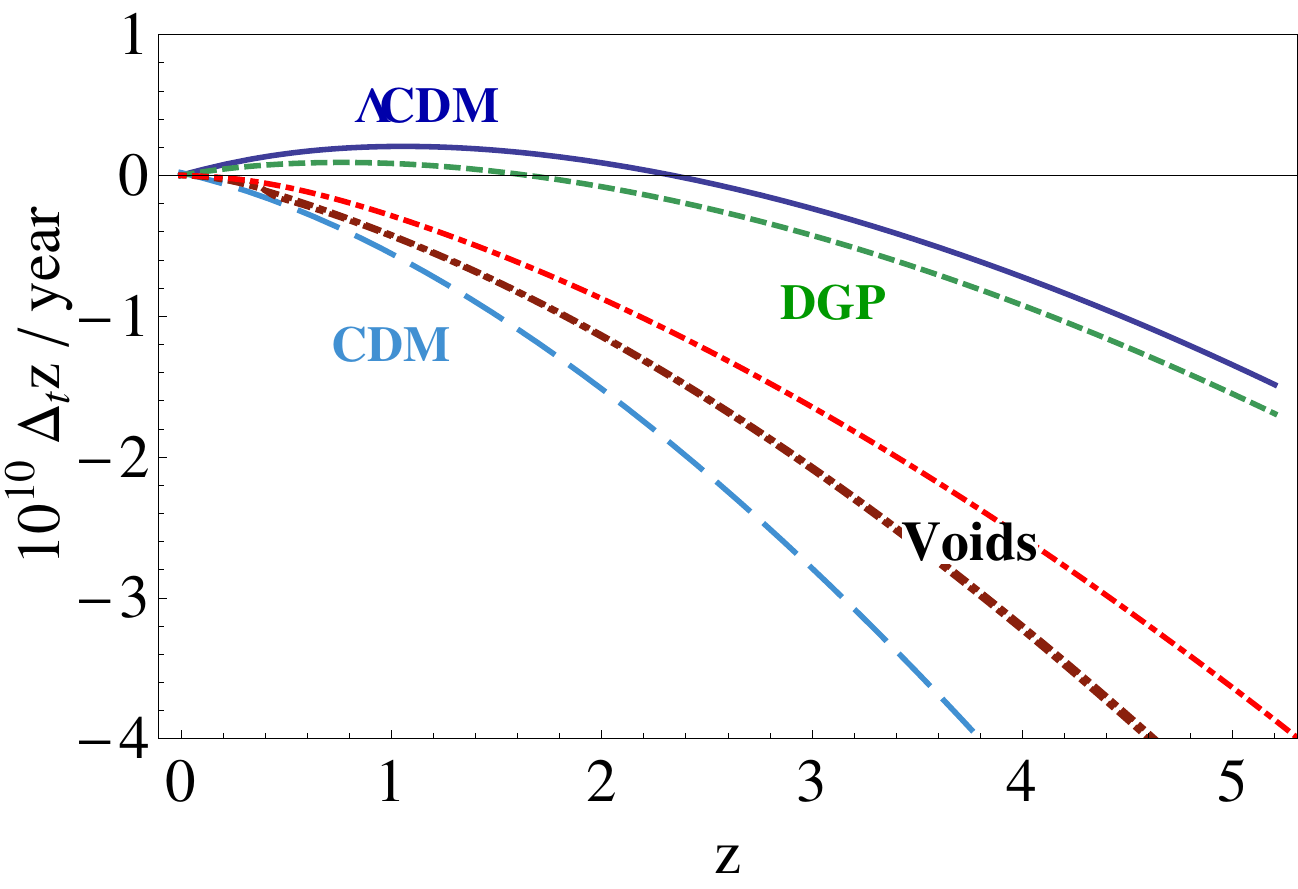}
\caption{The annual redshift drift for different models assuming an observer
at the center. The upper, blue solid lines represent the \lcdm\ model.
The green, dashed line corresponds to a self-accelerating DGP model
with $\Omega_{r_{c}}=0.13$. The dot-dashed lines stand for the 3
void models considered here: the dark brown (indistinguishable) lines
are for Models I and II, while the red line just above correspond
to the cGBH model. The bottommost line corresponds to an universe
with only matter in a FRW metric (the CDM model). Note that the void
models predict a curve which is in between CDM and \lcdm\, but closer
to the former.}
\label{fig:zdot-per-year}
\end{figure}

Finally, figure~\ref{fig:zdot-per-year} illustrates the Sandage
effect as a function of redshift for \lcdm\, the DGP model~\cite{dgp}, the
old matter dominated model (CDM) and the 3 different void models here
considered. As could be expected, the void models predict a curve
which is in between CDM and \lcdm. Since the signal there is
closer to the CDM one, this makes for a potentially powerful probe
for distinguishing these dark-energy-like void models and \lcdm,
as we will see in detail in Section~\ref{sec:measure-zdot}. Note that our results are in qualitative agreement with the ones obtained in~\cite{yoo08}.

\section{Measuring the Redshift Drift with CODEX}

\label{sec:measure-zdot}

Recently, two high-precision spectrographs were proposed which could
in principle be used for measuring the redshift drift: the Cosmic
Dynamics Experiment (CODEX)~\cite{codex,liske08,cristiani07} at
the European Extremely Large Telescope~(E-ELT)~\cite{gilmozzi07}
and the Echelle Spectrograph for PREcision Super Stable Observations
(ESPRESSO)~\cite{espresso,liske08,cristiani07} at the Very Large
Telescope array (VLT). Although proposed later, ESPRESSO would serve
as a prototype implementation on the technology behind CODEX as part
of its feasibility studies and could be operational several years
before that experiment~\cite{liske08,cristiani07}.

The possibility of detecting the redshift drift with CODEX was analyzed
in a couple of papers~\cite{chm,balbi07,liske08,cristiani07}. In
particular, it was shown in~\cite{balbi07} that using reasonable
mission specifications for CODEX, a discrimination amongst many different
proposed dark energy models would not be possible in a time-frame
of less than 30~years. Here we will show that void models, on the
other hand, are much easier to tell apart through the Sandage effect
than other dark energy models. Using very similar mission specifications
for CODEX, we estimate that a $5\sigma$ detection/exclusion is possible
with less than 10 years of observation.

The achievable accuracy on $\sigma_{\Delta v}$ by the CODEX experiment
was estimated (through Monte Carlo simulations)~\cite{codex} to
be
\begin{equation}
    \sigma_{\Delta v}=1.35\left(\frac{\mbox{S/N}}{2370}\right)^{-1}\!\!\left(\frac{\nqso}{30}\right)^{-\frac{1}{2}} \!\!\left(\frac{1+\zqso}{5}\right)^{q}\mbox{cm/s},\label{eq:sigma-v}
\end{equation}
with
 \begin{equation}
    q\equiv-1.7\;\mbox{ for }\; z\le4\,,\quad\quad q\equiv-0.9\;\mbox{ for }\; z>4\,,
\end{equation}
where $S/N$ is the signal-to-noise ratio per pixel, $\nqso$ is
the total number of quasar spectra observed and $\zqso$ their redshift.
Note also that the error pre-factor 1.35 corresponds to using all
available absorption lines, including metal lines; using only \lya\ lines enlarges this pre-factor to 2~\cite{liske08}.

The signal-to-noise ratio per pixel was estimated in~\cite{liske08}
to be
\begin{equation}
    \frac{\mbox{S}}{\mbox{N}}=700\left[\frac{Z_{X}}{Z_{r}}\;10^{0.4(16-m_{X})}\; \left(\frac{D}{42\,\mbox{m}}\right)^{2}\;\frac{t_{{\rm int}}}{10\,\mbox{h}}\;\frac{\epsilon}{0.25}\right]^{\frac{1}{2}},\label{eq:StoN}
\end{equation}
where $Z_{X}$ and $m_{X}$ are the source zero point and apparent
magnitude in the ``X'' band and $D$, $t_{{\rm int}}$ and $\epsilon$
are the telescope diameter, total integration time and total efficiency
respectively. We assumed a pixel size of $0.0125$~{\AA} and a
central obscuration of the telescope's primary collecting area of
$10\%$~\cite{liske08}. Note that $D=42$~m corresponds to the
reference design for the E-ELT~\cite{gilmozzi07}.

The reason we quoted magnitudes in terms of an arbitrary ``X''
band is because one should use the magnitude of the bluest filter
that still lies entirely redwards of the quasar's \lya\ emission
line~\cite{liske08}. This means that for $\zqso<2.2$ one should
use the magnitude in the $g$-band; for $2.2<\zqso<3.47$ the one
in the $r$-band; for $3.47<\zqso<4.61$ the $i$-band; for $\zqso>4.61$
the $z$-band. A good estimate for $m_{X}$ can be achieved with the
SDSS DR7, selecting the brightest quasars in each redshift bin using
the appropriate band for such bin. Following~\cite{balbi07} we
will select 40 quasars in 5 redshift bins, centered at $z=\{2,\,2.75,\,3.5,\,4.25,\,5\}$,
all of the same redshift width of $0.75$. The corresponding bands
are, in order, $\{g,r,r,i,z\}$ (where the $i$-band could equally
be chosen for the middle bin). Doing so, one gets for the average
(amongst the 8 brightest quasars) apparent magnitude $m_{X}$ for
each bin the following: $m_{X}=\{15.45,\,16.54,\,16.40,\,17.51,\,18.33\}$.
Finally, we estimate the zero point magnitude ratio in each bin to
be~\cite{fornal07}: $Z_{X}/Z_{r}=\{1.01,\,1.00,\,1.00,\,0.98,\,0.93\}$.
The accuracy of this last estimate is however quite unimportant in what
follows.

One remark is in order before we proceed. In~\eqref{eq:sigma-v}
it was tacitly assumed that the observational strategy concentrates
all spectroscopic observations in the two endpoints of the interval
$\Delta t$ and that $t_{{\rm int}}\ll\Delta t$. First of all, in
order to obtain a good ($>2000$) $S/N$ with E-ELT, $t_{{\rm int}}$
is not negligible compared to $\Delta t$. Second, it has been claimed
in~\cite{liske08} that in principle it would be preferable to spread
the observations more evenly over $\Delta t$, although the same authors
conclude that the best strategy to minimize the errors would be to
concentrate as much as possible the telescope time in both the beginning
and ending of $\Delta t$. Either way, the error estimate~\eqref{eq:sigma-v}
is changed somewhat, but never by more than a factor 2. However, estimating
such a correction depends on the details of the observational strategy
and is beyond the scope of this work; therefore in what follows we
will neglect this possibility.

\begin{figure}[t!]
\includegraphics[width=8.8cm]{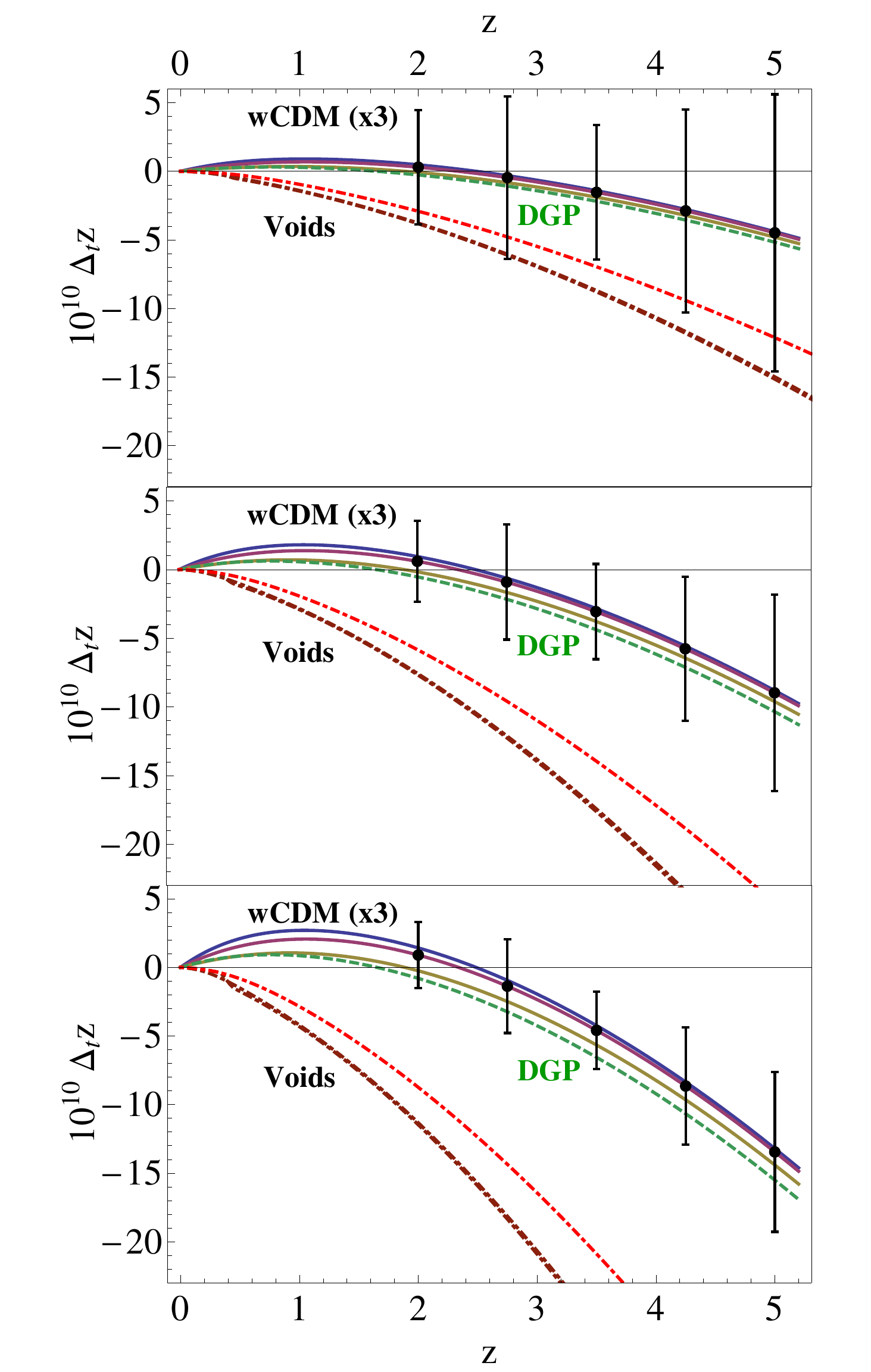}
\caption{Redshift drift for different dark energy models for a total mission
duration of 5 (top), 10 (middle) and 15 (bottom plot) years and CODEX
forecast error bars (the time-span between each measurement is 2/3
of that -- see text). In each plot, the upper 3, solid lines represent
wCDM models for $w=-1.25$ (uppermost), $w=-1$ (second) and $w=-0.75$
(third uppermost). The green, dashed line corresponds to a self-accelerating
DGP model with $\Omega_{r_{c}}=0.13$. The three bottommost, dot-dashed
lines stand for the 3 void models considered here: the dark brown
(indistinguishable) lines are for Models I and II, while the red line
just above correspond to the cGBH model. Note that a $4\sigma$ separation
between voids and $\Lambda$CDM can be achieved in a decade.}
\label{fig:zdot-ltb-codex}
\end{figure}

Hereafter we will therefore assume a compromise strategy: a three-period
observation, each of $\Delta t/3$ duration, and with observations
contained in the first and third periods. Doing so means that the
effective $\Delta t$ for the Sandage drift is $2\Delta t/3$.\footnote{Although they do not explicitly mention what
observational strategy they follow, it seems that the authors in~\cite{balbi07} in fact overestimated the redshift drift signal by a factor of 2 by assuming the total time interval of observation (in their proposal, 30~years) to coincide with the time interval in the redshift drift signal. The latter, for observations taken evenly over $\Delta t$ should in fact
be half the former (15~years in their case).} We will investigate three possible mission durations: 5, 10 and
15 years. It is important to note that a larger observational time-frame
allows not only for a larger redshift drift (which is linear in time)
but also for smaller error bars, as more photons are collected and,
therefore, a higher S/N (which increases as $\sqrt{\Delta t}$) can
be achieved. In other words, the {}``effective signal'' increases
with $\Delta t^{3/2}$ if one assumes a proportional telescope time
is maintained.

Figure~\ref{fig:zdot-ltb-codex} depicts the Sandage effect for different
dark energy models for three possible (complete observation) time-spans:
5, 10 and 15 years. Also plotted are the forecasted error bars obtainable
by CODEX at the E-ELT, the formula for which was discussed above.
Here we are assuming that the time spent observing each quasar is
the same, and this accounts for larger error bars at high redshift
due to the lower apparent magnitudes of the corresponding quasars
(see~\eqref{eq:StoN}). Another possible strategy would be to increase
the relative integration time for these sources in order to achieve
the same average signal-to-noise ratio at all redshift bins. Table~\ref{tab:zdot-codex}
contains the corresponding $\chi^{2}$ and $\sigma$-levels for 5,
10 and 15 years. As can be seen, void models could be detected/ruled
out at over $4\sigma$ with less than a decade of mission-time.

There is nothing special about the redshift binning proposed here. In fact, one could think about what would be the optimal redshift range for distinguishing between voids and \lcdm. By inspection of Figure~\ref{fig:zdot-ltb-codex}, it seems that the ``pivot redshift bin'' is the third one, centered around $z=3.5$. The reason is that this is the best compromise between the difference in the signal between the void and \lcdm\ (which increase with $z$) and quasar brightness (which decrease with $z$). In fact, if we only use the 8~QSOs in that bin we could improve the detection levels to $\{ 1.8\sigma, \,8.0\sigma,\, 15.7\sigma \}$ for Models I or II in $\{5, 10, 15\}$ years. However, this might not be desirable for two reasons: first, the pivot bins for other dark energy models are likely to be different; second, using more than a handful of quasars is important to wash out possible systematics, and actually for the SDSS catalog, using 40~QSOs all around $z=3.5$ \emph{decreases} the detection levels compared to the proposed binning (to $\{ 0.8\sigma, \,5.3\sigma,\, 11.0\sigma \}$ for the Models I or II) because we are then forced to use some not-so-bright quasars.

One very interesting aspect of using the Sandage effect to probe void
models is the fact it is model-independent to a good degree. In fact,
although in the cGBH model the signal is a little smaller, both Models
I and II here studied never differ by more than $0.1\sigma$ and except
for tiny differences close to the void edge (barely resolvable in
Figure~\ref{fig:zdot-ltb-codex}), they both predict the very same
redshift drift profile. These models should be good representatives
of this whole class of these dark-energy mimicking LTB void models. Model
I is very smooth, Model II represents an abrupt change between {}``inside''
and {}``outside'' the void and the cGBH model represents one of
the largest (over $2\unit{Gpc}$) voids in the literature.

\begin{table}[t]
\begin{tabular}{|c|c|c|c|}
    \hline
    \parbox[c][0.6cm]{0.5cm}{}\hspace{.5cm} \textbf{Model} \hspace{.5cm}  &  \textbf{ 5 years }   &  \textbf{ 10 years }   &  \textbf{ 15 years }  \tabularnewline
    \hline
    Models I / II  & \parbox[c][0.7cm]{0.5cm}{} $1.1\sigma$  & $6.2\sigma$  & $12.5\sigma$ \tabularnewline
    \cline{2-4}
    $ $  & \parbox[c][0.7cm]{0.5cm}{}$\chi^{2}=6.5$  & $\chi^{2}=52$  & $\chi^{2}=176$ \tabularnewline
    \hline
    cGBH Model  & \parbox[c][0.7cm]{0.5cm}{} .5$\sigma$  & 4.3$\sigma$  & $9.2\sigma$ \tabularnewline
    \cline{2-4}
    $ $  & \parbox[c][0.7cm]{0.5cm}{} $\chi^{2}=3.7$  & $\chi^{2}=30$  & $\chi^{2}=100$ \tabularnewline
    \hline
\end{tabular}
\caption{Estimated achievable confidence levels by the CODEX mission in 5,
10 and 15 years.}
\label{tab:zdot-codex}
\end{table}

In the next section we focus on the other real-time observable, cosmic
parallax.

\section{Measuring the Cosmic Parallax with Gaia}

\label{sec:measure-cp}

Distance measurements are one of the most fundamental challenges in
astronomy. The simplest and historically more important method to
measure cosmic distances relies on parallaxes, i.e., the apparent
change in position of an object relative to some reference frame generated
by a known displacement of the observer. In all astronomical applications
these displacements are small compared to the distances of the source:
the lunar parallax is around 1 degree; planetary parallaxes%
\footnote{Lunar and planetary parallaxes are measured from two different points
on the surface of the Earth, and therefore have a baseline limited
by our planet's diameter.} are $\lesssim$~11 arcsec; stellar parallax are $\lesssim$~1 arcsec;
galactic parallaxes\footnote{Stellar and galactic parallaxes are measured from two different positions
along the orbit of the Earth around the Sun, and therefore have a
maximum baseline of 2 AU.} are $\lesssim$~1~$\mu$as. Therefore measuring parallaxes of distant
sources require enough precision to detect tiny angular changes in
position. Even though observation of parallaxes on supergalactic scales
are daunting, of all (large) distance measurements they present the
least amount of systematics. This is the main reason why astrometry
has recently re-acquired an important role among the ground-based
and space-based planned missions. Measuring a possible apparent change
in the relative position of cosmological sources like quasars in any
anisotropic expansion scenario, dubbed in~\cite{quercellini09a}
cosmic parallax, is one of the next challenges for astrometry.

In particular, missions that perform \emph{global astrometry} over
the entire celestial sphere are preferred because: (i)~increasing
the number of measurement helps increasing the required accuracy;
(ii)~cosmic parallax is an all-sky effect, the multipole expansion
of which depend on (and therefore is a signature of) the underlying
anisotropic model. Such programmes measure the positions of objects
relative to other objects separated by a large angle on the sky, such
that they have a different parallactic effect. Therefore these missions
demands the capability to survey large and complete (flux-limited)
sample of objects. In ground-spaced programmes the observations are
typically done over a small field of view. In addition, the choice
of going to space offers the usual advantages of a stable thermal
environment, freedom from gravity and the atmosphere, and full sky
visibility. This factors enable the high-precision wide-angle astrometry
as implemented on missions such as Gaia~\cite{perryman01,bailer-jones05}
and SIM Lite Astrometric Observatory~\cite{goullioud08,simbook}.

Gaia is an European Space Agency (ESA) mission that will be launched
in 2012 with a nominal duration of 6 years. It marks a significant
step forward in astrometry, moving into the era of microarcsecond
astronomy and greatly extending Hipparcos' capabilities. The goal
is to achieve an astrometric accuracy (for the positional error $\sigma_{p}$)
between 10~$\mu$as (for sources with magnitude 15 on the $G$ band)
and 140~$\mu$as (for $G=20$)~\cite{perryman01} (although the final accuracy may be lower according to a revised estimate in~\cite{lindegren08}),
which should be compared to Hipparcos' 1000~$\mu$as astrometry and limiting magnitude
of 12. Gaia will also produce a full-sky map of roughly $0.5-1.0\;10^{6}$
quasars and $10^{9}$ stars down to its limiting magnitude of $G=20$,
whose positions will be determined (on average) with the above accuracy.
Direct optical observations of quasars is an important aspect of Gaia.
These will be observed in all of its 15 photometric bands at 100 epochs
from which the classes of quasars and their variability may be studied.
The relevance of measuring quasars is heightened due to the fact that
a fraction of them will be used to define the reference frame with
respect to which the positions of all other objects will be compared.

The observing strategy for Gaia (a drifting sky-scan) is not optimal for observing the CP, which would benefit from maximizing the time interval between quasar observations. However, even if the observational programme does not take into account the CP, in any case it constitutes at least a systematic that should not be ignored as it enlarges the astrometric error of any
global astrometry mission.

\begin{figure}[t!]
    \includegraphics[width=7.7cm]{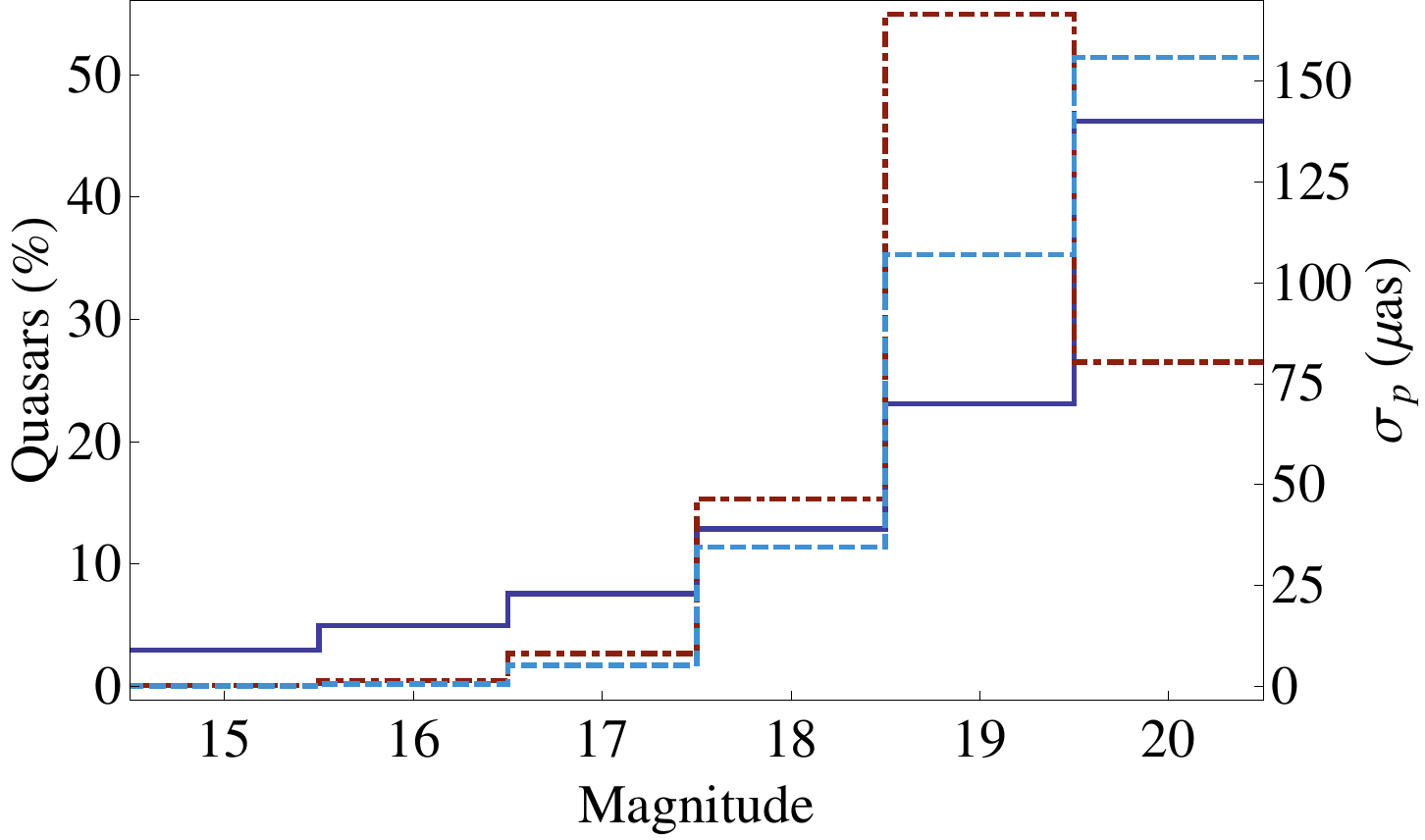}
    \caption{Target Gaia astrometric accuracy (dark, full lines) and projected
    quasar distribution (light, dashed) as a function of magnitude in
    the $G$ and $I$ band, respectively. Also plotted is the quasar distribution
    obtained using SDSS DR7 (red, dot-dashed), plotted against its own
    $G$ band magnitude. A simple weighted average gives the typical positional
    precision of Gaia on quasars: either $102\,\mu$as (projected) or
    $82\,\mu$as (SDSS-like distribution).}
    \label{fig:gaia-quasar-distrib}
\end{figure}

A rough estimate of the quasar distribution that Gaia is expected
to see comes from the observations made by the Sloan Digital Sky Survey
(SDSS)~\cite{sdss}. An earlier (pre-SDSS) but more adequate estimate
on this distribution was made in~\cite{gaiadraft} running a simulation
using Gaia's parameters, but using the $I$ instead of the $G$ band.
The mission's target astrometric accuracy as a function of magnitude
was derived in~\cite{perryman01}. Figure~\ref{fig:gaia-quasar-distrib}
depicts these forecasts as a function of the magnitude in the $G$
band. The plotted quasar distribution was obtained through the Sky
Server~\cite{skyserver} using the seventh data release (DR7) of
SDSS, which in the magnitude range of Gaia encompasses nearly $100000$
quasars. Combining these predictions allows us to estimate the average
positional precision of Gaia on quasars by taking a simple weighted
average: $102\,\mu$as (with the projection in~\cite{gaiadraft})
or $82\,\mu$as (SDSS-like distribution). Based on these predictions
we shall henceforth assume an average precision of $90\,\mu$as.

To compare our observations to Gaia we need to evaluate the average
$\Delta_{t}\gamma$ with $\Delta t=6$~years and $N$ sources. The
average angular separation of random points on a sphere is $\pi/2$
and thus the average of $\Delta_{t}\gamma$ can be estimated simply
as $\Delta_{t}\gamma(\theta=\pi/2)$. The final Gaia error $\sigma_{p}$
is obtained by best-fitting $2N$ independent coordinates from $N^{2}/2$
angular separation measures; the average positional error on the entire
sky will thus scale as $(2N)^{-1/2}$. The error scales therefore
as $\sigma_{p}/\sqrt{2\,\nqso}$. Since the CP signal increases linearly
with time, it is convenient to define
\begin{equation}
    \sqrt{\nqso}\,\left(\frac{\Delta t}{1\unit{year}}\right)\left(\frac{\sigma_{p}}{1\unit{\mu as}}\right)^{-1},\label{eq:FOM}
\end{equation}
which makes for a good figure-of-merit (FOM) for cosmic parallax
measurements. In the definition above, $\Delta t$ is the average
time interval between the two measurement epochs, and $\sigma_{p}$
is the average positional astrometric accuracy achieved \emph{in each
epoch}. With $\,\nqso=5\cdot10^{5}\,$ and $\,90\unit{\mu as}$, Gaia's
FOM is $\,39$. With a million sources, the FOM increases to around
$\,55$. In the Appendix~\ref{sec:sim} we show that, as a cosmic
parallax measuring mission, SIM Lite is less promising than
Gaia, boasting a FOM of only~$\,9$.

\begin{figure}[t!]
    \includegraphics[width=8.7cm]{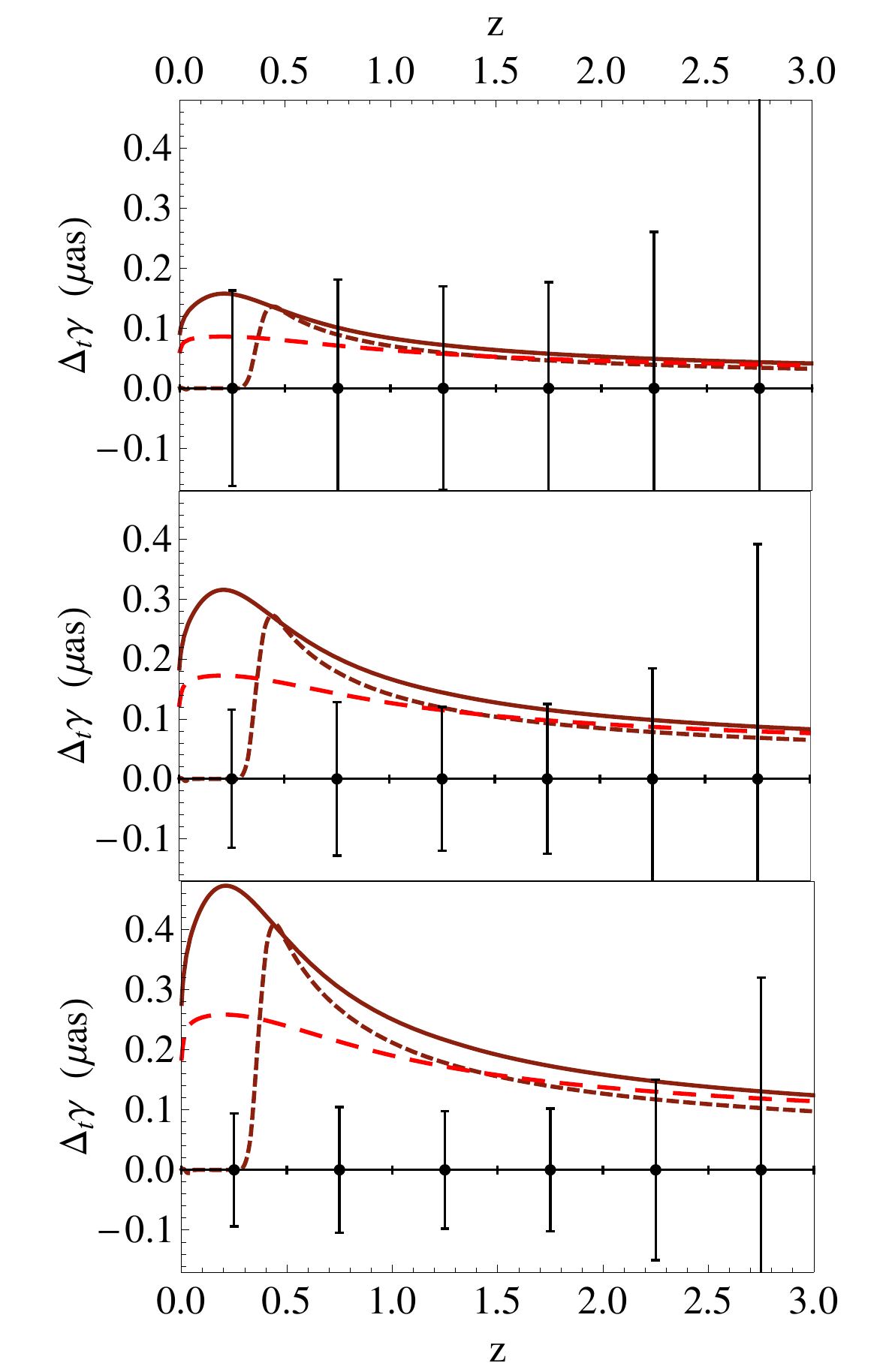}
    \caption{$\Delta_{t}\gamma$ for two sources at the same shell for both Models
    I (full) and II (dashed lines) and for a time span of 10 (top), 20
    (middle) and 30 (bottom plot) years, together with Gaia forecast statistical
    error bars. Although the nominal Gaia duration is only 6 years, a mission extension allow for
    smaller errors. Here we are not considering the two main systematics
    identified in the text. The lines correspond to the average cosmic
    parallax effect over the whole sky which is given by~\eqref{eq:average-cp}.
    Note that the CP quickly becomes the best probe of present anisotropy
    and, therefore, of the combination of distance and velocity towards
    the center of a void.}
    \label{fig:cp-and-gaia}
\end{figure}

Figure~\ref{fig:cp-and-gaia} illustrates the possibility of detecting
the cosmic parallax with Gaia for a possible, though arbitrary, redshift
binning. Depicted are $\,\Delta_{t}\gamma\,$ for two sources at the
same shell for both Models I (full) and II (dashed lines) and for
a time span of 10 (top), 20 (middle) and 30 (bottom plot) years, together
with Gaia forecast statistical error bars. The error bars are given
by
\begin{equation}
    \left<\sigma_{p}\right>/\sqrt{2\,\nqso}\label{eq:gaia-errors}
\end{equation}
in each bin, where $\left<\sigma_{p}\right>$ is Gaia's magnitude-averaged
precision on the corresponding bin. These errors correspond to the
previous nominal mission duration of 5 years and assume the SDSS-like quasar
distribution (see Figure~\ref{fig:gaia-quasar-distrib}). An extension
to 10 or more years allow smaller error bars and here too we can
approximate the errors to scale as $(\Delta t)^{-1/2}$. For $\, z>3$,
the error bars get much larger and the CP is quite small, so that
higher-$z$ bins do not add much. Here we are not considering the
two main source of systematics identified below. As in Figure~\ref{fig:dgammath-vs-z},
the lines correspond to a separation of $90^{\circ}$ in the sky between
the sources, which is the average separation between any two sources
in the sky. Table~\ref{tab:CP-Gaia} contains the corresponding $\chi^{2}$
and $\sigma$-levels.

Let us now come back to the matter of the fiducial off-center distance, raised in Section~\ref{sec:off-center}. We have so far assumed such a distance to be $30\unit{Mpc}$, which is the largest distance in agreement with the CMB dipole for an observer without peculiar velocity. Since the cosmic parallax signal is directly proportional to such a distance, one could also phrase the argument of detection in a different way. If we ignore the CMB dipole (and all other) dipolar-anisotropy constraints and leave the off-center distance as a free parameter, how well could Gaia constrain it? To estimate this one need only calculate, for a given number of mission years, what is the off-center distance that would produce a $1\sigma$ detection. Table~\ref{tab:CP-Gaia-off-center} summarizes the results for all 3 models in 6, 10, 20 and 30 years. Interestingly, although a Gaia-like mission requires around 20 years to reach the constraining level of the CMB dipole, already with 6 years it is an equivalent or even better probe of dipolar anisotropy in comparison to current supernovae datasets, which only limit such a distance to around 200-400~Mpc depending on the model~\cite{blomqvist09}.

\begin{table}[t]
\begin{tabular}{|c|c|c|c|}
    \hline\parbox[c][0.6cm]{0.5cm}{}\hspace{.5cm} \textbf{Model} \hspace{.5cm}  &  \textbf{ 10 years }   &  \textbf{ 20 years }   &  \textbf{ 30 years }  \tabularnewline
    \hline
    Model I  & \parbox[c][0.7cm]{0.5cm}{} $.05\sigma$  & $1.8\sigma$  & $4.9\sigma$ \tabularnewline
    \cline{2-4}
    $ $  & \parbox[c][0.7cm]{0.5cm}{}$\chi^{2}=1.4$  & $\chi^{2}=11$  & $\chi^{2}=39$ \tabularnewline
    \hline
    Model II  & \parbox[c][0.7cm]{0.5cm}{} $.003\sigma$  & $.5\sigma$  & $2.2\sigma$ \tabularnewline
    \cline{2-4}
    $ $  & \parbox[c][0.7cm]{0.5cm}{}$\chi^{2}=.5$  & $\chi^{2}=4.3$  & $\chi^{2}=19$ \tabularnewline
    \hline
    cGBH Model  & \parbox[c][0.7cm]{0.5cm}{} .005$\sigma$  & .6$\sigma$  & $2.6\sigma$ \tabularnewline
    \cline{2-4}
    $ $  & \parbox[c][0.7cm]{0.5cm}{}$\chi^{2}=.6$  & $\chi^{2}=5$  & $\chi^{2}=17$ \tabularnewline
    \hline
\end{tabular}
\caption{Estimated achievable confidence levels by the Gaia (or an extended Gaia-like) mission in 10, 20 and 30 years, in the limit where the two considered systematics are arbitrarily distinguished apart. For the Gaia's nominal duration of 6 years, detection levels are essentially zero.}
\label{tab:CP-Gaia}
\end{table}

\begin{table}[t]
\begin{tabular}{|c|c|c|c|c|}
    \hline\parbox[c][0.6cm]{0.5cm}{}\hspace{.5cm} \textbf{Model} \hspace{.5cm}  &  \textbf{6 years} &  \textbf{10 years}   &  \textbf{20 years}   &  \textbf{30 years}  \tabularnewline
    \hline
    Model I  & \parbox[c][0.7cm]{0.5cm}{} $143$  &$66$  & $23$  & $13$ \tabularnewline
    \hline
    Model II  & \parbox[c][0.7cm]{0.5cm}{} $235$  &$109$  & $39$  & $21$ \tabularnewline
    \hline
    cGBH Model  & \parbox[c][0.7cm]{0.5cm}{} $214$  &$99$  & $35$  & $19$ \tabularnewline
    \hline
\end{tabular}
\caption{Estimated off-center distance constraints (in Mpc) from the Gaia (or an extended
Gaia-like) mission in 6, 10, 20 and 30 years, in the limit where the two
considered systematics are arbitrarily distinguished apart.}
\label{tab:CP-Gaia-off-center}
\end{table}

Clearly, the Gaia mission with its nominal duration of 6 years cannot
detect the cosmic parallax in void models. For a longer mission duration,
however, detection (say, $3\sigma$) could be in principle achieved
with less than the 30 years estimated in Table~\ref{tab:CP-Gaia}.
The reason is twofold. First, earlier estimates for Gaia hinted to
the possibility of detecting up to 1 million quasars, which is twice
the value we are considering here. This extra data, if confirmed,
would amount to an extra $2\sigma$ to the detection levels in 30
years in any of the three models. Second, we only considered here
a simplified strategy of binning quasars in redshift, which amounts
to comparing the cosmic parallax of sources at same distances. But
in principle one should also compare quasars at different redshifts,
and this could lead to an average higher signal. Finally, one should
also take into account the $\phi$-coordinate in the distribution
of the quasars, and doing so should change the estimates somewhat.
We leave the last two points, however, for future work.

It is important to note that, although Gaia uses a fraction of the quasars to self-calibrate its inertial reference frame, these are only used to correct for rotations, which is a basically independent degree of freedom. In other words, all observed quasars can be used to reduce the errors statistically~\cite{lindegren}.

Two local effects induce spurious parallaxes (the observation of which are interesting on its own): one (of the order of 0.1~$\mu$as/year) is induced by our own peculiar velocity\footnote{Since the void is not expect to be moving with respect to the quasars frame, the peculiar velocity signal should be understood as one between our local group and the center of the void.} and the other (of the order of 4~$\mu$as/year~\cite{kovalevsky03}) by a changing aberration\footnote{Aberration is an optical distortion effect in the sky whenever observer and sources have nonzero relative velocities (see, for instance,~\cite{calvao05}).} due to the observers' acceleration. In astronomy in general (and cosmic parallax is no exception) a possible constant aberration is irrelevant. However, since the Sun is accelerating towards the center of the Milky Way, the resulting change in aberration does produce a competing signal which must be distinguished. This acceleration is the dominant competing effect, and even though the orbit around the galaxy is not circular, the extra yearly aberration due to this acceleration is given by the familiar centripetal acceleration formula~\cite{kovalevsky03} $a_{\sun}=V_{{\rm rot}}^{2}/R_{\sun}$. Current uncertainties on these two (sometimes called fundamental) parameters are around 5-10\%~\cite{vsun-rsun}, but radio astrometry at the Very Long Baseline Array (VLBA) might bring these down to~1\% within one decade~\cite{reid09}, which would imply around 3\% precision in $a_{\sun}$. Although this could in principle be used to predict and therefore subtract 97\% of this changing aberration, amounting to a residual signal of approximately 0.1~$\mu$as/year, such a procedure is not necessary: the best way to tell apart aberration effects from cosmic parallax is through their distinct redshift dependance (see Figure~\ref{fig:dgammath-vs-z}).

Both changing aberration and our own peculiar velocity produce a dipolar parallax signal, just like in LTB. However, as per our comments following~\eqref{eq:pecvel}, the peculiar velocity parallax decreases monotonically with the angular diameter distance (but not with redshift), while the aberration residual noise is \emph{independent of distance}. In contrast, the LTB signal has a characteristic non-trivial dependence on redshift: for the models investigated here it is either moderate (Model I) or vanishingly small (Model II) inside the void, large near the edge and decreasing at large distances (see Figure~\ref{fig:dgammath-vs-z}). It is therefore in principle possible to separate the cosmic signal from the (residual)
local ones, for instance estimating the local effects from sources inside the void. In fact, Milky Way stars form a gravitationally bound system and are not subject to cosmic parallax. They can therefore be used to self-calibrate Gaia and help separate the aberration-induced signal. A detailed calculation of the detection levels obtainable by Gaia requires not only taking these two systematics into account, but also a careful simulation of experimental settings (including possibly effects like source photocenter jitter and relativistic light deflection by solar system bodies) which is outside the scope
of this paper.

One final note regarding these systematics: more general (non-LTB) anisotropic models will not produce a simple dipole~\cite{quercellini09b,fontanini09} and their cosmic parallax can be more easily distinguished from local
effects.

\section{Conclusions}

In this paper we have presented two methods to map large scale inhomogeneity and late-time anisotropy: the redshift drift and cosmic parallax, respectively. Together, these real-time observables can fully reconstruct the 3D cosmic flow of distant sources. We forecasted the effect induced by a large void centered, or nearly centered, on the Milky Way, and in particular we have shown that the two effects can be detected with the E-ELT and with Gaia or an enhanced version of Gaia. The two effects add to the limited number of tests that can be employed to distinguish a LTB void from an accelerating FRW universe, possibly eliminating an exotic alternative explanation to dark energy.


In LTB void models, the Sandage effect turns out to be mostly sensitive to the scale of the void, but not to other particular void properties like steepness of the transition. The CP, on the other hand, being basically an anisotropy probe is mostly sensitive to the off-center distance, and in fact should be zero for on-center observers. It also depends somewhat on the particular void profile, specially for $z \lesssim 0.5$. Nevertheless, although one can guess this low-$z$ (``inside'' the void) CP behavior for pathological cases like the very abrupt Model II (where it is zero, as per Figures~\ref{fig:dgammath-vs-z} and~\ref{fig:cp-and-gaia}), it is not obvious how exactly all the void parameters enter into the final effect.

It turns out that the best hope to attain a clear-cut discrimination between LTB and FRW is with the redshift drift effect, since the LTB expansion is always decelerated. We find that a 4$\sigma$ separation can be achieved with E-ELT in less than 10 years, much before the same experiment will be able to distinguish between competing models of dark energy. A Gaia-like mission, on the other hand, can only achieve a reasonable detection of a void-induced cosmic parallax in the course of 30 years.

Nevertheless, cosmic parallax remains an important tool and in fact one of the most promising way to probe general late-time cosmological anisotropy, as already discussed in~\cite{quercellini09a,quercellini09b}. In particular, even if it only lasts 6 years Gaia should constrain late-time anisotropies similarly to current supernovae catalogs, but in an independent way. Also,
in \lcdm\ it can be used to measure our own peculiar velocity with respect to the quasar reference frame and consequently to the CMB, therefore providing a new and promising way to break the degeneracy between the intrinsic CMB dipole and our own peculiar velocity. We are currently investigating this possibility and results will be published in a subsequent paper.

Direct kinematic tests as redshift drift and cosmic parallax are conceptually the simplest probe of expansion and of anisotropy since their interpretation do not rely on calibration of standard candles/rulers nor depend on evolutionary or selection effects (as for galaxy ages and number counts). The fact that in both CP and redshift drift the ``effective signal'' increases as $\left( \Delta t \right)^{3/2}$ shows that these new real-time cosmology effects can become some of the most effective long-term dark energy probes. For the same reason, the Sandage and cosmic parallax effects have also the potential to become the best inhomogeneity and (late-time) anisotropy tests, respectively. Combined, they will form an important direct test of the FRW metric.

Although the odds of Gaia having fuel to last 10 or more years are small, one can consider Gaia as making a first sub-miliarcsecond astrometric sky-map, which could be confronted with any future global-astrometry mission. Since any proper motion signal increases linearly with time, any future mission with a global astrometric accuracy \emph{at least} as good as Gaia can be used to detect the CP (or any other kind of late-time anisotropy) signal. In between missions, however, the effective signal grows only linearly in $\Delta t$.

It's really exciting that two great tools like Gaia and E-ELT are
becoming reality just now when we begin to realize the importance
of extremely precise astrometric and spectroscopic measurements for
cosmology.

\appendix

\section{Numerical Nuances}

\label{sec:numerical-nuances}

The intrinsically smallness of both the cosmic parallax and Sandage
effects demand a carefully constructed numerical code to correctly
compute either. As stated before, a straightforward calculation of
$\Delta_{t}\gamma$ per year requires evaluating $\xi_{a1}$ and $\xi_{a2}$
with at least 15 orders of precision (as the CP is of order $0.2\unit{\mu as/year}\sim10^{-13}$~rad/year).

It is possible to alleviate this by exploiting the linearity of $\Delta_{t}\gamma$
in $\Delta t$ and scaling up the system. In fact, we confirmed that
such linearity held at least up to $\Delta t=10^{6}$ years, so that
this was the value used in~\cite{quercellini09a} to compute the
CP, dividing in the end the result by $10^{6}$ to get the parallax-per-year.
However, even for such an enlarged time span, a CP estimation still
require an end-of-calculation precision of 9 digits; for the stated
algorithm, which involves solving 5 coupled differential equations
many times and comparing the results, this is not possible using standard
double-precision techniques.

The first method we resorted to used a simplification for the metric.
In the limit $|\alpha(r)\beta(r)/R(t,r)\ll1|$ (which always holds
in the models we investigated), the metric $R(t,r)$ can be written
explicitly~\cite{chung06} without resort to the parameter~$\eta$.
This allowed us to further exploit arbitrary-precision numerical routines
such as the ones found in Mathematica$^{\copyright}$ to carry on
our computations with a precision higher than the regular machine-precision
(16 digits of precision). Surprisingly, even though the metric approximation
is very reasonable, the results obtained were not consistent. This
is probably due to the fact that second derivatives appear in~\eqref{eq:geodesics}
(a good approximation to a function might not be so for its derivatives)
and also to the fact that  the stated algorithm is very sensitive
to any small deviations to the geodesics' paths.

Therefore we dropped the approximation in~\cite{chung06} in favor
of another one: setting $R_{{\rm lss}}$ to zero in~\eqref{eq:alpha}-\eqref{eq:beta}.
This has negligible impact on the metric for $z\lesssim10$. Doing
so allows us to invert~\eqref{eq:sol-beta-t} and obtain the function~$\eta\big[2\beta(r)^{3/2}\, t\,/\,\alpha(r)\big]$
and its first 2 derivatives using Mathematica's arbitrary precision
routines, thus computing the metric to a high-enough precision in
order to obtain consistent results. Since going above machine precision
slows down any code exponentially we must also be careful not to set
the target precision too high. Over different parts of the algorithm
we had to work with in between 20 and 30-digit precision.

Even when using high-precision techniques, numerical noise became
unstable whenever $\beta(r)$ became too close to zero (as can be
easily seen through~\eqref{eq:sol-R}-\eqref{eq:sol-beta-t}), so
we adopted a slightly modified version of~\eqref{eq:beta}: \begin{equation}
\beta(r)=\left(H_{\perp,0}^{{\rm out}}\right)^{2}r^{2}\,\frac{\Delta\alpha}{2}\left(1.001-\tanh\frac{r-r_{{\rm vo}}}{2\Delta r}\right)\!,\label{eq:beta-mod}\end{equation}
 where the only change was on the factor before the $\tanh$ from
$1$ to $1.001$. Does this affect the CP signal? We tested this for
both Models I and II by putting a higher factor of either $1.01$
or $1.05$ and found out that there is no change in any results, so
we assume the same should hold in the limit where this factor goes
to unity.

\section{Measuring the Cosmic Parallax with SIM Lite}

\label{sec:sim}

The SIM Lite Astrometric Observatory (a smaller, cost effective version
of the formerly known SIM PlanetQuest)~\cite{goullioud08,simbook}
is being developed by the Jet Propulsion Laboratory under contract
with NASA and has a target launch-date for around $2016$. Like Gaia,
it is also a astrometry-centered mission with a $5$-year nominal
duration, but one with different observational strategy and scientific
goals. One of its main objectives is the search for Earth-sized extrasolar
planets and therefore instead of pursuing a global astrometric measurement
it will focus on specific regions of the sky. In these narrow regions,
SIM Lite can achieve a higher precision compared to Gaia: $1\,\mu$as
with a single measurement and $4\,\mu$as for the global astrometric
grid. Nevertheless as we will show below, SIM Lite is less adequate
than Gaia for measuring the cosmic parallax, mostly due to the small
amount of time devoted to extragalactic observations. In fact, current
proposals call for an observation of only 50 quasars, devoting only
1.5\% of the mission duration for that purpose.

How does SIM Lite compare with respect to Gaia in measuring the cosmic
parallax? As discussed in Section~\ref{sec:measure-cp}, the precision
of such measurement scales as $\,\sigma_{p}/\sqrt{\nqso}$. For Gaia,
as shown, we estimate $\,\sigma_{p}=90\,\mu\mbox{as}\,$ and at least
$\,\nqso=500000$; for SIM Lite, $\,\sigma_{p}\approx4\unit{\mu as}\,$
and $\,\nqso=50\,$ (a selected sample with magnitude in the R band
less than $16.5$~\cite{simbook}). Therefore the CP figure-of-merit
(see Section~\ref{sec:measure-cp}) of SIM Lite is $9$, which is
over 4 times smaller than Gaia's FOM (which is $39$). Nevertheless,
SIM Lite only allocates 1.5\% of its mission time to observing quasars.
One could therefore question how much better could a similar instrument
do in observing the CP if it allocated 100\% of its time for that
purpose. A first estimate would then give $\,\nqso\approx3000\,$
(a realistic number, as SDSS DR7 contains a little over 1000 quasars
with $\, R<16.5$), and in this case such a mission would have around
double the precision (i.e., FOM) of Gaia. Since clearly Gaia's CP-measuring
capabilities could also be enhanced on a similar way by allocating
more integration-time for quasars, it remains the most promising current
proposal for that.

\section*{Acknowledgments}

The authors would like to thank Claudia Quercellini, Ulrich Bastian, Lennart Lindegren, Marie-Noëlle Célérier, Tomi Koivisto,  Sergei Klioner and the anonymous referee for fruitful discussions and/or suggestions. L. A. acknowledges financial contribution from contract ASI-INAF I/064/08/0.

Funding for the SDSS and SDSS-II has been provided by the Alfred P.
Sloan Foundation, the Participating Institutions, the National Science
Foundation, the U.S. Department of Energy, the National Aeronautics
and Space Administration, the Japanese Monbukagakusho, the Max Planck
Society, and the Higher Education Funding Council for England. The
SDSS Web Site is http://www.sdss.org/.

The SDSS is managed by the Astrophysical Research Consortium for the
Participating Institutions. The Participating Institutions are the
American Museum of Natural History, Astrophysical Institute Potsdam,
University of Basel, University of Cambridge, Case Western Reserve
University, University of Chicago, Drexel University, Fermilab, the
Institute for Advanced Study, the Japan Participation Group, Johns
Hopkins University, the Joint Institute for Nuclear Astrophysics,
the Kavli Institute for Particle Astrophysics and Cosmology, the Korean
Scientist Group, the Chinese Academy of Sciences (LAMOST), Los Alamos
National Laboratory, the Max-Planck-Institute for Astronomy (MPIA),
the Max-Planck-Institute for Astrophysics (MPA), New Mexico State
University, Ohio State University, University of Pittsburgh, University
of Portsmouth, Princeton University, the United States Naval Observatory,
and the University of Washington.

\end{document}